\documentclass[smallcondensed]{article}   
\usepackage[utf8]{inputenc}
\usepackage{cases}
\usepackage{mathptmx}
\usepackage{amssymb}
\usepackage{color, graphics, graphicx,epic,eepic}
\usepackage{natbib}
\usepackage{apalike}
\begin{document}
\providecommand\mathbb[1]{\ensuremath{\mathsf{#1}}}
\renewcommand{\theequation}{\thesection.\arabic{equation}}
\newcommand\BbbE{\mathbb{E}} 
\hbadness=10000
\vbadness=10000
\newcommand*{\rttensor}[1]{\bar{\bar{#1}}}
\title{Advection-diffusion in porous media with low scale separation:
modelling via higher-order asymptotic homogenisation}
\author{Pascale Royer}

\date{}
\maketitle
\begin{abstract}
Asymptotic multiple scale homogenisation allows to determine the effective behaviour of a porous medium 
by starting from the pore-scale description, when there is a large separation between the pore-scale and the macroscopic scale. 
When
the scale ratio is “small but not too small,” the standard
approach based on first-order homogenisation may
break down since additional terms need to be taken into
account in order to obtain an accurate picture of the overall response of the medium.
The effect of low scale separation can be obtained 
 by exploiting higher order equations in the asymptotic homogenisation procedure.
 The aim of the present study is to investigate higher-order terms up to the third order of the advective-diffusive model 
to describe advection-diffusion in a macroscopically homogeneous porous medium 
at low scale separation.
 The main result of the study is that the low separation
 of scales induces dispersion effects. In particular, 
 the second-order model is similar to the most currently used phenomenological model of dispersion: it is characterised by a dispersion tensor which
 can be decomposed
 into a purely diffusive component and a mechanical dispersion part, whilst this property is not verified in the homogenised dispersion model
 (obtained at higher P\'eclet number). The third-order description contains second and third concentration gradient terms, 
with a fourth order tensor of diffusion and with a third-order and an additional second-order tensors of dispersion.
The analysis of the macroscopic  fluxes shows that the second and the third order macroscopic fluxes are distinct from the volume averages of the corresponding local fluxes
and allows to determine  expressions of the non-local effects.
\end{abstract}
%
\section{Introduction}
\label{intro}
Most studies in the theory of flow and transport in porous media are based on the exploitation
of the continuum theory implying that the original heterogeneous
medium behaves like a homogeneous one characterised by macroscopic fluid flow and transport equations 
with certain effective properties. 
Such an approach requires that the condition of separation of scales be fulfilled:
the microscopic size $l$ of heterogeneities must be
essentially smaller than the macroscopic characteristic length $L$ : 
$l \ll L$. In this definition, length
$L$ represents either the size of the whole sample,
or a macroscopic characteristic length of the phenomenon, which means that
the condition of separation of scales must be fulfilled geometrically and also and with respect to loading conditions.

The multiple-scale asymptotic homogenisation method which can be traced to \cite{San80},  \cite{Ben78}, and
\cite{Bak89} can be used as a systematic tool of averaging so as to derive such continuum models:
first-order models obtained by asymptotic homogenisation are thus accurate
for media with large scale separation between the pore scale and the macroscale.
But when
the ratio $l/L$ is ``small but not too small``, microstructural scale effects may occur which result in specific non-local phenomena.
Then, the  “local action”  assumption of classical continuum mechanics, which postulates that
the current state of the medium  at a given point is only affected by its
immediate neighbours and  that there are no physical mechanisms that produce
action at a distance, is no longer satisfied.
Consequently, additional terms need to be taken into
account in order to obtain an accurate picture of the overall response of the medium,
which
cannot be predicted in the frame of first-order homogenisation  theory.
Thus, the study of so-called higher order or non-local effects in
the overall behaviour of heterogeneous media is motivated by the need to account for 
the scale effects  observed in the behaviour of multiple-scale heterogeneous
media where the scales are separated widely but not ``too widely'', and
these scale effects can be systematically analysed by considering  higher-order correctors
in the asymptotic homogenisation method.

Mathematical aspects of higher-order homogenisation have been 
developed in \cite{Smy00}, \cite{Che04}.
The role of higher order terms has
been investigated for heat conduction in heterogeneous materials in \cite{Bou95}
and  for elastic composite materials subjected to 
static loading in \cite{Gam89}, \cite{Bou96}.
In these studies, it is shown that the
heterogeneity of the medium  causes non-local
effects on a macrolevel: instead of the homogenised equilibrium equations of
continuum mechanics, new equilibrium equations are obtained  that involve 
higher-order spatial derivatives and thus represent the influence of the
microstructural heterogeneity on the macroscopic behaviour of the material.
In dynamic problems,  application of higher-order homogenisation
provides a long-wave approach valid in the low-frequency range \citep{Bou93, Fish01,Bak05,Chen01,And08}.
In \cite{Bou93}, it is demonstrated that higher-order terms successively introduce
effects of polarization, dispersion and attenuation.

Transport in porous media with low scale separation 
has thus far received relatively little attention. However, two important works on fluid flow have been performed.
In \cite{Goy97, Goy99}, the authors investigate the permeability in a
dendritic mushy zone, which is generally a nonhomogeneous porous structure. They make use of the volume averaging method to obtain corrector
terms to Darcy’s law. In \cite{Aur05}, the validity of Darcy's law is investigated by higher-order
asymptotic homogenisation up to the third order. 

The focus of the present study is on solute transport by advection-diffusion in porous media with low scale separation, which can occur in the two following situations \citep{Aur05}:
i) when large gradients of concentration
are applied to macroscopically homogeneous porous media; ii) when the porous medium is macroscopically heterogeneous and the macroscopic characteristic length $L$ associated to the macroscopic heterogeneities is not “very” large compared to the characteristic length $l$ of the pores. The scope of the present work is to derive  higher-order homogenised models of advection-diffusion in macroscopically homogeneous porous media and is therefore aimed at describing the situations where large concentration gradients are applied. This may for example happen during soil-column experiments, where soil samples are necessarily limited in size and are subjected to large concentration gradients, especially at early stages of the tests. In these situations the macroscopic characteristic length $L \approx C/\mid \vec\nabla C  \mid $ associated to this gradient of concentration is not “very” large compared
to $l$ \citep{Aur97}. 
Homogenisation of convection-diffusion equations on the pore scale leads to three macroscopic transport models,
accordingly to the order of magnitude of the P\'eclet number \citep{AurAdler95}: i) a diffusion model; ii) an advection-diffusion model; iii)
an advection-dispersion model.
Whilst the first two models are first-order models, the dispersive model requires to account for the 
first corrector.
The purpose of the present work is to derive the second and third order homogenised models in the case where 
the model of advection diffusion is obtained at the first order.
 
The  paper is organised as follows. Section \ref{Transp_models} presents the existing phenomenological and homogenised macro-models and their properties for describing
solute transport in rigid porous media. The input transport problem is formulated in Section \ref{prob_statement}: the medium geometry is described in \S\ref{geometry} and
the pore-scale  governing equations for fluid flow and solute transport are then presented and nondimmensionalised in \S\ref{gov-eqs}.
The results from \cite{Aur05} for higher-order homogenisation up to the third order of the fluid flow equations, and which are required for the developments
that follow, are briefly summarised in Section \ref{hom_fluid_flow}. Section \ref{hom_solute_transport} is devoted to higher-order homogenisation up to the third order
of solute transport equations in the advective-diffusive macro-regime. 
The physical meaning of the volume averages of local fluxes which arise with the homogenisation procedure is analysed in Section \ref{macro_fluxes} and 
the writing of the second and third order homogenised
models in terms of the macroscopic fluxes provides expressions of the non-local effects.
Finally, Section \ref{conclu}, 
presents a summary of the main theoretical results contained in this work and highlights  conclusive remarks.
\section{About phenomenological and homogenised models of solute transport in porous media}
\label{Transp_models}
\subsection{Phenomenological macro-models}
Let consider a rigid porous medium saturated by an incompressible Newtonian fluid.
When the fluid is at rest, transient solute transport within the medium is described by the model of diffusion:
\begin{equation}
 \phi\displaystyle\frac{\partial C}{\partial t}-\overrightarrow{\nabla}\cdot ({\bar{\bar{D}}}^{\hbox{\tiny eff}}\overrightarrow{\nabla} C) = 0,
 \label{pheno-diff}
\end{equation} 
in which $\phi$ denotes the porosity, $C$ represents the concentration and ${\bar{\bar{D}}}^{\hbox{\tiny eff}}$ is the tensor of effective diffusive.
When the fluid is in motion, solute transport may  either be described by the model of advection-diffusion
\begin{equation}
 \phi\displaystyle\frac{\partial C}{\partial t}-\overrightarrow{\nabla}\cdot ({\bar{\bar{D}}}^{\hbox{\tiny eff}}\overrightarrow{\nabla} C - C \overrightarrow{V}) = 0,
  \label{pheno-advec-diff}
\end{equation}
or by the model of advection-dispersion
\begin{equation}
 \phi\displaystyle\frac{\partial C}{\partial t}-\overrightarrow{\nabla}\cdot ({\bar{\bar{D}}}^{\hbox{\tiny disp}}\overrightarrow{\nabla} C - C \overrightarrow{V}) = 0.
 \label{pheno-model_disp}
\end{equation}
In both models, $\overrightarrow{V}$ denotes the macroscopic fluid velocity and verifies:
\begin{numcases}{}
 \overrightarrow{V}= - \frac{\bar{\bar{K}}}{\mu}\overrightarrow{\nabla} P,& (Darcy's law)\label{DL}\\
 \overrightarrow{\nabla} \cdot \overrightarrow{V} = 0,
\end{numcases}
where 
$\rttensor{K}$ 
denotes the tensor of permeability, $\mu$ is the fluid viscosity and $P$ represents the fluid pressure. For the sake of simplicity, gravity is neglected in Eq. (\ref{DL}).
Tensor ${\bar{\bar{D}}}^{\hbox{\tiny disp}}$ in model Eq. (\ref{pheno-model_disp}) is the tensor of hydrodynamic dispersion: it depends on the fluid velocity.
In the most currently used model of dispersion \citep{Bear72,BearBach90}, the tensor of dispersion is decomposed into the sum of a diffusive term and a term of mechanical dispersion
which depends on  the fluid velocity:
\begin{equation}
{\bar{\bar{D}}}^{\hbox{\tiny disp}}=
 {\bar{\bar{D}}}^{\hbox{\tiny eff}} +
 {\bar{\bar{D}}}^{\hbox{\tiny mech disp}}.
 \label{pheno-tensor_disp}
\end{equation}
Whilst the regime of advection-diffusion is rarely mentioned in the geosciences literature, 
it is of particular relevance for modelling electro-chemio-mechanical coupling
in swelling porous media \citep{MoyMur06a}. Advection-diffusion is furthermore the usual transport regime observed in biological tissues \citep{Beck13,Amb06,Swi10,Lem13}.
\subsection{Homogenised models}
%
\label{Transp_hom_models}
Homogenisation of the convection-diffusion equations on the pore scale allows to find the three above-mentioned transport regimes \citep{AurAdler95}
and to give their res\-pective range of validity
by means of the order of magnitude of the P\'eclet number
\begin{equation}
 {\mathbb P}e=\displaystyle\frac{v_c L}{D_c},
\end{equation}
where $L$ denotes the characteristic macroscopic length, and where $v_c$ and $D_c$ are characteristic values of the local fluid velocity and of the coefficient of molecular diffusion. 
The results of \citet{AurAdler95} are the following:
\begin{numcases}{}
 {\mathbb P}e\leq{\mathcal O}(\varepsilon):& Regime of diffusion\nonumber\\
 {\mathbb P}e={\mathcal O}(\varepsilon^0)& Regime of advection-diffusion\nonumber\\
 {\mathbb P}e={\mathcal O}(\varepsilon^{-1})& Regime of advection-dispersion\nonumber\\
{\mathbb P}e\geq{\mathcal O}(\varepsilon^{-2})& No continuum macro-model\nonumber,
\end{numcases}
where $\varepsilon=l/L $, with $l$ being the pore-scale characteristic length, is the small parameter of the asymptotic homogenisation method and where
a parameter $\mathbb{P}$ is said to be of order $\varepsilon^p $, $\mathbb{P}={\mathcal O}(\varepsilon^p) $, when
\begin{equation}
 \varepsilon^{p+1}\ll \mathbb{P}\ll \varepsilon^{p-1}.
 \label{rule_OdG}
\end{equation}
The homogenised models of diffusion and of advection-diffusion are first-order models
and are rigorously identical  to models Eqs. (\ref{pheno-diff})-(\ref{pheno-advec-diff}).
On the other hand however, the homogenised model of advection-dispersion is different from the classical phenomenological model Eq. (\ref{pheno-model_disp}).
It is a second-order model, which in particular implies that Darcy's law is no longer valid \citep{Aur05}.
Furthermore, the homogenised tensor of dispersion does not verify relationship Eq. (\ref{pheno-tensor_disp}) and is not symmetric \citep{AurAdler95, Aur10}.
At high P\'eclet number, ${\mathbb P}e\geq{\mathcal O}(\varepsilon^{-2})$, the problem becomes dependent upon the macroscopic boundary-conditions. Consequently, there exists no  continuum macro-model to
describe solute transport within this regime.
\section{Problem statement for homogenisation of solute transport within the advective-diffusive regime}
\label{prob_statement}
\setcounter{equation}{0}
\subsection{Geometry}
\begin{figure}[ht!!]
    \begin{center}
    \includegraphics[width=10cm]{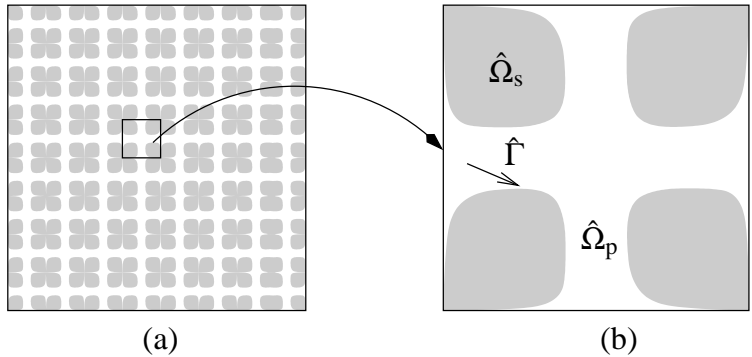}
    \end{center}
  \caption{\label{fig_milieu}{\it\small Periodic porous medium : (a) Macroscopic sample ; (b) Periodic cell $\hat\mathrm\Omega$.}}
    \end{figure}

%
\label{geometry}
Consider a rigid porous medium with connected pores. We assume it to
be periodic with period $\hat\mathrm\Omega$. The fluid
occupies the pores $\hat\mathrm\Omega_\mathrm p$, and $\hat\mathrm\Gamma$ represents the surface of the solid matrix $\hat\mathrm\Omega_\mathrm s$.
We denote as $\hat l$ and $\hat L$ the characteristic length
of the pores and the macroscopic length (Fig. \ref{fig_milieu}).
We assume the scales to be separated and we define
\begin{equation}
 \varepsilon = \displaystyle\frac{\hat l}{\hat L}\ll 1.
\end{equation}
Using the two characteristic lengths, $\hat l$ and $\hat L$, two dimensionless space variables are defined
\begin{eqnarray}
 \overrightarrow{y} = \displaystyle\frac{\overrightarrow{\hat X}}{\hat l}\hspace{0,5cm}\hbox{which describes variations on the microscopic scale},
 \\
 \overrightarrow {x}=\displaystyle\frac{\overrightarrow{\hat X}}{\hat L}\hspace{0,5cm}\hbox{which describes variations on the macroscopic scale,}
\end{eqnarray}
where ${\overrightarrow {\hat X}}$ is the physical spatial variable.
Invoking the differentiation rule of multiple var\-iables, the gradient operator with respect to ${\overrightarrow{\hat X}}$  is written as
\begin{equation}
 \overrightarrow{\nabla}_{\hat X}=\displaystyle\frac{1}{l}\overrightarrow{\nabla}_y+\displaystyle\frac{1}{L}\overrightarrow{\nabla}_x,
 \label{dim_grad_op}
\end{equation}
where $\overrightarrow{\nabla}_y$ and $\overrightarrow{\nabla}_x$ are the gradient operators with respect to $\overrightarrow{y}$ and $\overrightarrow{x}$,
respectively.
%
\subsection{Governing equations on the pore scale and estimates}
%
\label{gov-eqs}
The pores  are saturated with a viscous,
incompressible and Newtonian fluid containing a low concentration of solute $\hat c$.
The fluid
is in slow steady-state isothermal flow, so that the solute is transported by diffusion
and convection. 
%
\subsubsection{Fluid flow}
%
The equations governing velocity $\overrightarrow{\hat v}$ and pressure $\hat p$ 
of an incompressible viscous fluid of viscosity $\hat\mu$ in slow steady-state flow
within the pores are the following:\\
\\- Stokes equation
\begin{equation}
  \hat\mu\Delta_X\overrightarrow{\hat v} - \overrightarrow{\nabla}_{\hat X} \hat p=\overrightarrow 0
 \hspace{0,5cm}\hbox{within $\hat\Omega_{\mathrm p}$},
 \label{F_Eq1_dim}
\end{equation}
- the conservation of mass
\begin{equation}
  \overrightarrow{\nabla}_{\hat X}\cdot\overrightarrow{\hat v} = 0
 \hspace{0,5cm}\hbox{within $\hat\mathrm\Omega_{\mathrm p}$,}
 \label{F_Eq2_dim}
\end{equation}
- the no-slip condition
\begin{equation}
  \overrightarrow{\hat v} = \overrightarrow{0}\hspace{0,5cm}\hbox{over $\hat\mathrm\Gamma$.}
 \label{F_Eq3_dim}
\end{equation}
%
\subsubsection{Solute transport}
The transport of solute by diffusion-convection in the pore domain is described by conservation of mass
\begin{equation}
 \displaystyle\frac{\partial \hat c}{\partial \hat t} - \overrightarrow{\nabla}_{\hat X}\cdot(\hat D_0\overrightarrow{\nabla}_{\hat X}\hat c - \hat c
 \overrightarrow{\hat v}) = 0
 \hspace{0,5cm}\hbox{within $\hat\mathrm\Omega_{\mathrm p},$}
 \label{CD_Eq2_dim}
\end{equation}
and the no-flux boundary condition
\begin{equation}
 (\hat D_0\overrightarrow{\nabla}_{\hat X}\hat c - \hat c \overrightarrow{\hat v})\cdot\overrightarrow{n} =(\hat D_0\overrightarrow{\nabla}_{\hat X}\hat c)
 \cdot\overrightarrow{n} = 0
 \hspace{0,5cm}\hbox{over $\hat\mathrm\Gamma$,}
 \label{CD_Eq3_dim}
\end{equation}
where $\hat c$ is the solute concentration (mass of solute per unit volume of fluid), 
t is the time, $ \hat D_0$ denotes the coefficient of molecular diffusion and $\overrightarrow{n}$ is
the unit vector giving the normal to $\hat\mathrm\Gamma$ exterior to $\hat\mathrm\Omega_\mathrm p$.
%
\subsubsection{Nondimensionalisation
 and estimates}
%
Introducing into Eqs. (\ref{F_Eq1_dim})-(\ref{CD_Eq3_dim})
$$
\begin{array}{lll}
\overrightarrow{\nabla}_{\hat X}=1/L\ \overrightarrow{\nabla}, &\Delta_{\hat X}=1/L^2\ \Delta,&\vspace{0,2cm}\\
\hat t= t_\mathrm c\ t,&\ \partial/\partial  \hat t=1/t_\mathrm c\ \partial/\partial t,&\vspace{0,2cm}\\
\overrightarrow{\hat v}=v_\mathrm c\ \overrightarrow{v},&\ \hat p = p_\mathrm c\ p,&\ \hat c = c_\mathrm c\ c,\vspace{0,2cm}\\
\hat\mu=\mu_\mathrm c\ \mu,& \hat D_0=D_{\mathrm c}\ D_0,&\vspace{0,2cm}\\
 \end{array}
$$
where quantities with subscript $\mathrm c$ denote characteristic quantities,
we can write the microscopic description in dimensionless form as
\begin{numcases}{}
 \mathbb{F}\ \mu\Delta\overrightarrow{v} - \overrightarrow{\nabla} p=\overrightarrow 0
 \hspace{0,5cm}\hbox{within $\Omega_\mathrm p $,}
 \label{F_Eq1}\\
 \overrightarrow{\nabla}\cdot\overrightarrow{v} = 0
 \hspace{0,5cm}\hbox{within $\Omega_\mathrm p $,}
 \label{F_Eq2}\\
\mathbb{N}\ \displaystyle\frac{\partial c}{\partial t} - \overrightarrow{\nabla}\cdot(D_0\overrightarrow{\nabla}  c- \mathbb{P}e\ c \overrightarrow{v}) = 0
\hspace{0,5cm}\hbox{within $\Omega_\mathrm p $,}
\label{CD_Eq2}\\
\overrightarrow{v} = \overrightarrow{0}\hspace{0,5cm}\hbox{over $\Gamma$,}
 \label{F_Eq3}\\
 (D_0\overrightarrow{\nabla}  c) \cdot \overrightarrow n = 0\hspace{0,5cm}\hbox{over $\Gamma$,}
\label{CD_Eq3}
\end{numcases}
with
$$
\mathbb{F}=\displaystyle\frac{\mu_\mathrm c v_\mathrm c}{L p_\mathrm c};\hspace{0,5cm}
\mathbb{N}=\displaystyle\frac{ L^2}{t_\mathrm c D_\mathrm c};\hspace{0,5cm}
\mathbb{P}\mathrm{e} = \displaystyle\frac{v_\mathrm c L}{D_\mathrm c}.
$$
In the above writing,  
the dimensionless counterpart of any dimensional quantity 
$\hat\Psi$ is $\Psi=\hat\Psi/\Psi_\mathrm c$. 
In particular, the characteristic time $t_\mathrm c$ is the time over which we intend to describe the
solute transport: it is the characteristic time of the observation.
We have arbitrarily chosen the macroscopic length $\hat L$  as the reference length for normalising the gradient operator. 
Consequently, according to Eq. (\ref{dim_grad_op}),
the corresponding dimensionless gradient operator reads
\begin{equation}
 \overrightarrow{\nabla}=L\overrightarrow{\nabla}_{\hat X}=\varepsilon^{-1}\overrightarrow{\nabla}_y+\overrightarrow\nabla_x.
 \label{adim_grad_op}
\end{equation}
We may now estimate the three dimensionless parameters, $ \mathbb{F}$, $\mathbb{N}$
and the Péclet number $\mathbb{P}e$, with respect to powers of the small parameter $\varepsilon$
and for this purpose we  shall apply the rule defined by Eq. (\ref{rule_OdG}).
Parameter $\mathbb{F}$, which arises from Stokes equation Eq. (\ref{F_Eq1}), is the ratio of the viscous term to the pressure gradient. 
We shall consider the case where 
homogenisation of Stokes equations  leads to Darcy's law on the sample scale. As shown in \cite{Aur91}, this happens when
the local flow is balanced by a macroscopic pressure gradient, which in an order-of-magnitude sense reads
\begin{equation}
 \displaystyle\frac{\mu_\mathrm c v_\mathrm c}{l^2}={\mathcal O}\left(\displaystyle\frac{p_\mathrm c}{L}  \right),
\end{equation}
and yields
\begin{equation}
  \mathbb{F}=\displaystyle\frac{\mu_\mathrm c v_\mathrm c}{L p_\mathrm c}={\mathcal O}(\varepsilon^2).
  \label{OdG_F}
\end{equation}
The order-of-magnitude of the Péclet number $\mathbb{P}\mathrm{e}$  characterises the regime of solute transport. 
Indeed, it is the ratio of characteristic times
of diffusion and convection
\begin{equation}
 \mathbb{P}\mathrm{e}=
 \displaystyle\frac{t^{\scriptscriptstyle\mathrm{diff}}}{t^{\scriptscriptstyle\mathrm{conv}}},
\end{equation}
where
\begin{numcases}{}
  t_{\scriptscriptstyle}^{\scriptscriptstyle\mathrm{diff}}=\displaystyle\frac{L^2}{D_\mathrm c}
 &{(macroscopic characteristic time of diffusion)},\\
  t_{\scriptscriptstyle}^{\scriptscriptstyle\mathrm{conv}}=\displaystyle\frac{L}{v_\mathrm c}&
  (macroscopic characteristic time of convection).
\end{numcases}
We consider 
\begin{equation}
 \mathbb{P}\mathrm{e}  =\displaystyle\frac{v_\mathrm c L}{D_\mathrm c}= {\mathcal O}(\varepsilon^0),
 \label{Odg_Pe}
\end{equation}
which leads to the homogenised advective-diffusive model at the first order ({\em Cf.} \S\ref{Transp_hom_models}).
The dimensionless number $\mathbb{N}$ is such that:
\begin{equation}
 \mathbb{N}=\displaystyle\frac{t_{\scriptscriptstyle\mathrm L}^{\scriptscriptstyle\mathrm{diff}}}{t_\mathrm c }.
\end{equation}
Since  $ \mathbb{P}\mathrm{e}  = {\mathcal O}(\varepsilon^0)$ means that $t^{\scriptscriptstyle\mathrm{diff}}=
t^{\scriptscriptstyle\mathrm{conv}}$ , we take $t_\mathrm c=t^{\scriptscriptstyle\mathrm{diff}}=
t^{\scriptscriptstyle\mathrm{conv}}$, which yields
\begin{equation}
  \mathbb{N} = {\mathcal O}(\varepsilon^0).
 \label{OdG_N}
\end{equation}
Note that taking
$t_\mathrm c=t^{\scriptscriptstyle\mathrm{diff}}=
t^{\scriptscriptstyle\mathrm{conv}}$ ensures a macroscopic transient regime, while $ t_\mathrm c>t^{\scriptscriptstyle\mathrm{diff}}$ would lead to a macroscopic steady-state regime
and that when $t_\mathrm c<t^{\scriptscriptstyle\mathrm{diff}}$, the transport mechanism is not sufficiently developed for its evolution be  described by means of a continuum model.
\section{Higher-order homogenisation of fluid flow}
%
\label{hom_fluid_flow}
\setcounter{equation}{0}
Homogenisation of the fluid flow equations has been performed up to the third order in \cite{Aur05}.
Equations Eqs. (\ref{F_Eq1})-(\ref{F_Eq3}) are considered with Eq. (\ref{OdG_F}), which leads to 
the following set of flow equations
\begin{numcases}{}
 \varepsilon^2\mu\Delta\overrightarrow{v} - \overrightarrow{\nabla} p=\overrightarrow{0}
 \hspace{0,5cm}\hbox{within $\Omega_\mathrm p$,}
 \\
 \overrightarrow{\nabla}\cdot\overrightarrow{v}= 0
 \hspace{0,5cm}\hbox{within $\Omega_\mathrm p$,}
 \label{F_eq2}\\
 \overrightarrow{v}=\overrightarrow{0}
 \hspace{0,5cm}\hbox{over $\Gamma$,}
 \label{F_eq3}
\end{numcases}
where
\begin{equation}
 \overrightarrow{\nabla} = \varepsilon^{-1}\overrightarrow{\nabla}_y+\overrightarrow{\nabla}_x.
 \label{space_deriv_adim}
\end{equation}
The homogenisation procedure consists in looking for the pressure and the velocity
in the form of asymptotic expansions in powers of $\varepsilon$ \citep{Ben78, San80}:
\begin{numcases}{}
 p(\overrightarrow{y}, \overrightarrow{x})  = p^0(\overrightarrow{y}, \overrightarrow{x})  + \varepsilon p^1(\overrightarrow{y}, \overrightarrow{x}) 
 + \varepsilon p^2(\overrightarrow{y}, \overrightarrow{x}) +...&\nonumber
 \\ 
 \overrightarrow{v}(\overrightarrow{y}, \overrightarrow{x})  = \overrightarrow{v}^0(\overrightarrow{y}, \overrightarrow{x}) 
 +\varepsilon \overrightarrow{v}^1(\overrightarrow{y}, \overrightarrow{x}) 
 +\varepsilon^2 \overrightarrow{v}^2(\overrightarrow{y}, \overrightarrow{x}) +...\nonumber
\end{numcases}
For a macroscopically homogeneous
medium, the results can be summarised as follows
\begin{equation}
 \displaystyle\frac{\partial}{\partial x_i}(< v_i^n >)=0 \hspace{0,5cm}(n=0, 1, 2),
 \label{F_macro_bal}
\end{equation}
with
\begin{numcases}{}
 <v_i^0>=-\displaystyle\frac{K_{ij}}{\mu}\displaystyle\frac{\partial p^0}{\partial x_j},
 \\
  <v_i^1>=-\displaystyle\frac{N_{ijk}}{\mu}\displaystyle\frac{\partial^2 p^0}{\partial x_j\partial x_k}-\displaystyle\frac{K_{ij}}{\mu}\displaystyle\frac{\partial \bar p^1}{\partial x_j},\label{av_v1}
 \\
 <v_i^2>=-\displaystyle\frac{P_{ijkl}}{\mu}\displaystyle\frac{\partial^3 P^0}{\partial x_j\partial x_k\partial x_l}
 -\displaystyle\frac{N_{ijk}}{\mu}\displaystyle\frac{\partial^2 \bar p^1}{\partial x_j\partial x_k}
 -\displaystyle\frac{K_{ij}}{\mu}\displaystyle\frac{\partial \bar p^2}{\partial x_j},\label{av_v2}
\end{numcases}
where $<.>$ denotes the volume average and is defined by
\begin{equation}
 <.> = \displaystyle\frac{1}{\mid\Omega\mid}\int_{\Omega_\mathrm p}\ .\ d\Omega.
\end{equation}
The third order tensor $N_{ijk}$ is symmetric with respect to its last two indices and antisymmetric with respect to its first two indices.
Then, since $N_{ijk}$  is symmetrical with respect to its  last two indices, it is equal to zero when the medium is isotropic.\\
\\Functions $p^0$, $p^1$ and $p^2$ are such that
\begin{numcases}{}
 p^0 =p^0 (\overrightarrow{x}),
 \\
 p^1 = -a_j(\overrightarrow{y})\displaystyle\frac{\partial p ^0}{\partial x_j}+\bar p^1 (\overrightarrow{x}),
\label{p1}\\
 p^2= -d_{jk}(\overrightarrow{y})\displaystyle\frac{\partial^2 p^0}{\partial x_j\partial x_k}-
 a_j(\overrightarrow{y})\displaystyle\frac{\partial \bar p^1}{\partial x_j}+\bar p^2 (\overrightarrow{x}).\label{p2}
\end{numcases}
Note that functions $\bar p^1$ and $\bar p^2$, which appear in Eqs. (\ref{av_v1}) and (\ref{av_v2}), are particular solutions involved in the definitions of $p^1$ and $p^2$, Eqs. (\ref{p1}) and (\ref{p2}), respectively.
Combining Eq. (\ref{F_macro_bal}) with the averaged velocities, the second-gradient terms vanish as a result of the antisymmetry of $N_{ijk}$. 
Thus, the following flow descriptions are obtained
\begin{numcases}{}
\hbox{(First order)}\hspace{0,5cm} \displaystyle\frac{\partial}{\partial x_i}(K_{ij}\displaystyle\frac{\partial p^0}{\partial x_j})=0,
\\
 \hbox{(Second order)}\hspace{0,5cm} \displaystyle\frac{\partial}{\partial x_i}(K_{ij}\displaystyle\frac{\partial \bar p^1}{\partial x_j})=0,
 \\
 \hbox{(Third order)}\hspace{0,5cm} \displaystyle\frac{\partial}{\partial x_i}(P_{ijkl}\displaystyle\frac{\partial^3 p^0}{\partial x_j\partial x_k\partial x_l}
 +K_{ij}\displaystyle\frac{\partial \bar p^2}{\partial x_j})=0.
\end{numcases}
%
\section{Higher-order homogenisation of solute transport in the advective-diffusive regime}
%
\label{hom_solute_transport}
\setcounter{equation}{0}
\subsection{Local dimensionless description}
%
\label{Transp_loc_desc}
We consider Eq. (\ref{CD_Eq2}) with  estimates Eq. (\ref{Odg_Pe}) and  Eq. (\ref{OdG_N}), and boundary conditions
Eqs. (\ref{F_Eq3})-(\ref{CD_Eq3}).
This leads to the following  set of equations:
\begin{numcases}{}
 \displaystyle\frac{\partial c}{\partial t}-\overrightarrow{\nabla}\cdot (D_0\overrightarrow{\nabla}  c-  c \overrightarrow{v})=0
 \hspace{0,5cm}\hbox{within $\Omega_\mathrm p $},\label{CD_Eq2_adim}&\\
\overrightarrow{v} = \overrightarrow{0}
\hspace{0,5cm}\hbox{over $\Gamma$}\label{F_Eq3_adim},&\\
 (D_0\overrightarrow{\nabla}  c)\cdot\overrightarrow{n}= 0
 \hspace{0,5cm}\hbox{over $\Gamma$}\label{CD_Eq4_adim}.&
\end{numcases}
We look for solutions to the unknowns $c$ and $\overrightarrow{v}$  of the form:
\begin{numcases}{}
 c(\overrightarrow{y}, \overrightarrow{x}) =c^0(\overrightarrow{y}, \overrightarrow{x}) + \varepsilon c^1(\overrightarrow{y}, \overrightarrow{x}) 
 + \varepsilon^2 c^2(\overrightarrow{y}, \overrightarrow{x})   +...&\nonumber
 \\
 \overrightarrow{v}(\overrightarrow{y}, \overrightarrow{x}) = \overrightarrow{v}^0(\overrightarrow{y}, \overrightarrow{x}) 
 +\varepsilon \overrightarrow{v}^1 (\overrightarrow{y}, \overrightarrow{x})  +\varepsilon^2 \overrightarrow{v}^2 (\overrightarrow{y}, \overrightarrow{x}) +...
 &\nonumber
\end{numcases}
where functions $c^n(\overrightarrow{y}, \overrightarrow{x})$ and  $\overrightarrow{v}^n(\overrightarrow{y}, \overrightarrow{x})$ are $\Omega$-periodic in $\overrightarrow{y}$. Furthermore, because of the two spatial
variables $\overrightarrow{x}$ and $\overrightarrow{y} = \varepsilon^{-1}\overrightarrow{x}$, the spatial derivation takes
the  form Eq. (\ref{space_deriv_adim}). The homogenisation technique involves the introduction of these expansions 
into the dimensionless equations Eqs. (\ref{CD_Eq2_adim})-(\ref{CD_Eq4_adim}) 
and the identification of the powers of $\varepsilon$.
%
\subsection{First-order homogenisation}
%
\label{Transp_1storder_hom}
\subsubsection{Boundary value problem for $c^0$}
%
At the first order, the boundary value problem  Eqs. (\ref{CD_Eq2_adim})-(\ref{CD_Eq4_adim}) leads to:
\begin{numcases}{}
 \displaystyle\frac{\partial}{\partial y_i}\left(D_0 \displaystyle\frac{\partial c^0}{\partial y_i} \right) = 0
 \hspace{0,5cm}\hbox{in $\Omega_\mathrm p$},&\\
 D_0\displaystyle\frac{\partial c^0}{\partial y_i}n_i = 0
 \hspace{0,5cm}\hbox{over $\Gamma$},&\\
 c^0: \hbox{periodic in $\overrightarrow{y}$,}
\end{numcases}
from which it is clear that the concentration $c^0$ is constant over the period
\begin{equation}
 c^0 = c^0(\overrightarrow{x}, t).
 \label{c0}
 \end{equation}
 %
\subsubsection{Boundary value problem for $c^1$}
%
We now consider the second order of Eqs. (\ref{CD_Eq2_adim})-(\ref{CD_Eq4_adim}). Then, noticing that (see Eq. (\ref{F_eq2})) 
\begin{equation}
 \displaystyle\frac{\partial v_i^0}{\partial y_i}=0,
 \label{divyv0}
\end{equation}
we obtain the following boundary value problem for $c^1$:
\begin{numcases}{}
  \displaystyle\frac{\partial}{\partial y_i}\left[D_0 (\displaystyle\frac{\partial c^1}{\partial y_i}+\displaystyle\frac{\partial c^0}{\partial x_i})  \right]=0
 \hspace{0,5cm}\hbox{within $\Omega_\mathrm{p}, $}\label{CD_Eq5_2bis}&\\
  \left[
 D_0(\displaystyle\frac{\partial c^1}{\partial y_i}+\displaystyle\frac{\partial c^0}{\partial x_i})  
 \right]n_i=0\hspace{0,5cm}\hbox{over $\Gamma,$}\label{CD_Eq7_2bis}&\\
 c^1: \hbox{periodic in $\overrightarrow{y}$.}
\end{numcases}
By virtue of linearity, the solution reads:
\begin{equation}
 c^1=\chi_j(\overrightarrow{y})\displaystyle\frac{\partial c^0}{\partial x_j} + \bar c^1(\overrightarrow{x}, t),
 \label{D_defc1}
\end{equation}
where $\bar c^1(\overrightarrow{x}, t)$ is an arbitrary function. 
The exact definition of the vector $\overrightarrow{\chi}$ is reported in Appendix \ref{1st_ord_localprob}.
Note that, to render the solution  unique, we impose that $\overrightarrow{\chi}$ is average to zero
\citep{Ben78,San80,Mei10}:
\begin{equation}
 <\overrightarrow{\chi}> = \displaystyle\frac{1}{\mid\Omega\mid}\int_{\Omega_\mathrm{p}}\ \overrightarrow{\chi}\ d\Omega = \overrightarrow{0}.
\end{equation}
Note further that, since we are considering a macroscopically homogeneous medium, $\overrightarrow{\chi}$ 
doesn't depend on variable $\overrightarrow{x}$: $\overrightarrow{\chi}=\overrightarrow{\chi}(\overrightarrow{y})$.
%
\subsubsection{Derivation of the first-order macroscopic description}
%
Let consider the boundary value problem Eqs. (\ref{CD_Eq2_adim})-(\ref{CD_Eq4_adim}) at the third order: 
\begin{numcases}{}
 \displaystyle\frac{\partial c^0}{\partial t}-\displaystyle\frac{\partial}{\partial y_i}\left[D_0(\displaystyle\frac{\partial c^2}{\partial y_i}+\displaystyle\frac{\partial c^1}{\partial x_i}) 
 -c^0v_i^1 - c^1v_i^0\right] 
 &\nonumber\\
 -\displaystyle\frac{\partial}{\partial x_i}\left[D_0 (\displaystyle\frac{\partial c^1}{\partial y_i}+\displaystyle\frac{\partial c^0}{\partial x_i})-c^0v_i^0\right]=0
  \hspace{0,5cm}\hbox{within $\Omega_\mathrm{p}, $}
  \label{CD_Eq5_3}&\\
  v_i^0=v_i^1=0\hspace{0,5cm}\hbox{over $\Gamma,$}\label{F_Eq3_3}\\
  \left[D_0(\displaystyle\frac{\partial c^2}{\partial y_i}+\displaystyle\frac{\partial c^1}{\partial x_i}) 
\right] n_i=0
 \hspace{0,5cm}\hbox{over $\Gamma.$}\label{CD_Eq7_3}&
\end{numcases}
The homogenisation procedure consists now in integrating Eq. (\ref{CD_Eq5_3}) over $\Omega_\mathrm p$. 
This leads to the so called compatibility condition, which is a necessary and
sufficient condition for the existence of
solutions. Furthermore, it represents the first-order macroscopic description.
Invoking Gauss' theorem, the integration yields:
\begin{equation}
\begin{array}{l}
 \displaystyle\displaystyle\frac{1}{\mid\Omega\mid}\displaystyle\int_{\Omega_p}\displaystyle\frac{\partial c^0}{\partial t}\ d\Omega 
 -  \displaystyle\displaystyle\frac{1}{\mid \Omega \mid}\displaystyle\int_{\delta\Omega_p}\left[D_0(\displaystyle\frac{\partial c^2}{\partial y_i}+\displaystyle\frac{\partial c^1}{\partial x_i}) 
 -c^0v_i^1 - c^1v_i^0\right]n_i\ dS\vspace{0,2cm}\\
 - \displaystyle\displaystyle\frac{1}{\mid\Omega\mid}\displaystyle \displaystyle\int_{\Omega_p}
  \displaystyle\displaystyle\frac{\partial}{\partial x_i}\left[D_0 ( \displaystyle\displaystyle\frac{\partial c^1}{\partial y_i}
  + \displaystyle\displaystyle\frac{\partial c^0}{\partial x_i})-c^0v_i^0\right]\ d\Omega = 0,
 \end{array}
 \label{D_int_Eq2_3bis}
\end{equation}
where $\delta\Omega_{\mathrm p}=\Gamma \cup (\delta\Omega\cap\delta\Omega_\mathrm p)$ denotes the bounding surface of $\Omega_{\mathrm p}$.
The second term of Eq. (\ref{D_int_Eq2_3bis}) is thus the sum of two surface integrals and it actually cancels out:
the integral over the surface $\Gamma$ vanishes because of boundary 
conditions Eqs. (\ref{F_Eq3_3})-(\ref{CD_Eq7_3}), while the integral over the cell boundary, $\delta\Omega\cap\delta\Omega_\mathrm p $, vanishes by periodicity.
Hence, Eq. (\ref{D_int_Eq2_3bis}) reduces to
\begin{equation}
 \phi \displaystyle\frac{\partial c^0}{\partial t}
 -\displaystyle\frac{\partial}{\partial x_i}<D_0(\displaystyle\frac{\partial c^1}{\partial y_i}+\displaystyle\frac{\partial c^0}{\partial x_i})-c^0v_i^0 >=0,
  \label{D_1st_ord_macro_0}
\end{equation}
where
\begin{equation}
 \phi = \displaystyle\frac{\mid \Omega_\mathrm{p} \mid}{\mid \Omega\mid}
\end{equation}
denotes the porosity.
Using Eq. (\ref{D_defc1}), we can write:
\begin{equation}
 \displaystyle\frac{\partial c^1}{\partial y_i}+\displaystyle\frac{\partial c^0}{\partial x_i}=\gamma_{ij}^0\displaystyle\frac{\partial c^0}{\partial x_j},
 \label{gradyc1_plus_gradxc0}
\end{equation}
where
\begin{equation}
 \gamma_{ij}^0=\displaystyle\frac{\partial \chi_j}{\partial y_i}+\delta_{ij}.
 \label{def_gamma0}
\end{equation}
Taking  Eq. (\ref{F_macro_bal}) into account, Eq. (\ref{D_1st_ord_macro_0}) can be rewritten as follows:
\begin{equation}
 \phi\displaystyle\frac{\partial c^0}{\partial t} - \displaystyle\frac{\partial}{\partial x_i}(D_{ij}\displaystyle\frac{\partial c^0}{\partial x_j})
 + <v_i^0 >\displaystyle\frac{\partial c^0}{\partial x_i}=0,
 \label{CD_1st_macro_1}
\end{equation}
where
\begin{equation}
D_{ij} =\displaystyle\frac{1}{\mid\Omega\mid}\int_{\Omega_\mathrm{p}}\ D_0(\displaystyle\frac{\partial\chi_j}{\partial y_i}+\delta_{ij})\ d\Omega
=\displaystyle\frac{1}{\mid\Omega\mid}\int_{\Omega_\mathrm{p}}\ D_0\gamma_{ij}^0\ d\Omega
\end{equation}
is the tensor of effective diffusion. It can be shown that the second-order tensor  $D_{ij}$ is positive and  symmetric ({\em Cf.} Appendix \ref{prop_D}). \\
Defining the first-order macroscopic concentration and average fluid velocity by
\begin{numcases}{}
 <c> = <c^0> + {\mathcal O}(\varepsilon < c>),&\\
 <\overrightarrow{v}>=<\overrightarrow{v}^0>+{\mathcal O}(\varepsilon <\overrightarrow{v}>),&
\end{numcases}
the first-order macroscopic description thus reads
 \begin{equation}
 \phi\displaystyle\frac{\partial <c>}{\partial t} - \displaystyle\frac{\partial}{\partial x_i}(D_{ij}\displaystyle\frac{\partial <c>}{\partial x_j})
 + <v_i >\displaystyle\frac{\partial <c>}{\partial x_i}={\mathcal O}(\varepsilon \phi \displaystyle\frac{\partial <c >}{\partial t}).
 \label{CD_1st_macro_adim}
\end{equation}
In dimensional variables, it becomes
\begin{equation}
 \begin{array}{l}
 \displaystyle\phi\displaystyle\frac{\partial <\hat c>}{\partial \hat t} 
 - \displaystyle\displaystyle\frac{\partial}{\partial \hat X_i}(\hat D_{ij}^{\hbox{\tiny diff}}\displaystyle\frac{\partial <\hat c>}{\partial \hat X_j})
 + <\hat v_i >\displaystyle\displaystyle\frac{\partial <\hat c>}{\partial \hat X_i}
  =\displaystyle{\mathcal O}(\varepsilon \phi \displaystyle\frac{\partial <\hat c >}{\partial \hat t}),
 \end{array}
\end{equation}
where
\begin{equation}
 \hat D_{ij}^{\hbox{\tiny diff}} = D_\mathrm c\  D_{ij}
\end{equation}
is the tensor of effective diffusion.
The fluid velocity verifies ({\em Cf.} Section \ref{hom_fluid_flow}):
\begin{numcases}{}
 <\hat v_i >= - \displaystyle\frac{\hat K_{ij}^{\hbox{\tiny eff}}}{\hat\mu}\displaystyle\frac{\partial <\hat p>}{\partial \hat X_j}+ {\mathcal O}(\varepsilon <\hat v_i >),&\\
 \displaystyle\frac{\partial <\hat v_i >}{\partial \hat X_i}={\mathcal O}(\varepsilon \displaystyle\frac{\partial <\hat v_i >}{\partial \hat X_i}).&
\end{numcases}
The first-order behaviour is thus described by the classical advection-diffusion transport equation, in which the fluid velocity verifies Darcy's law.
\subsection{Second-order homogenisation}
%
\label{Transp_2ndorder_hom}
\subsubsection{Boundary value problem for $c^2$}
%
The third-order boundary value given by Eqs. (\ref{CD_Eq5_3})-(\ref{CD_Eq7_3}), can be transformed ({\em Cf.} Appendix \ref{BVP_c2}) so as to obtain the following boundary value problem
for $c^2$:
\begin{numcases}{}
 \frac{\partial}{\partial y_i}\left[D_0 (\frac{\partial c^2}{\partial y_i}+\chi_j\frac{\partial^2 c^0}{\partial x_i\partial x_j}
  +\frac{\partial \bar c^1}{\partial x_i})  \right]=&\nonumber\\
  (\frac{1}{\phi}D_{ij} - D_0\gamma_{ij}^0)\frac{\partial^2 c^0}{\partial x_i\partial x_j}
  +(v_i^0\gamma_{ij}^0 -\frac{1}{\phi}<v_j^0>)\frac{\partial c^0}{\partial x_j}
   \hspace{0,5cm}\hbox{within $\Omega_\mathrm{p} $,}\label{CD_eqdefc2}&\\
   \left[D_0 (\frac{\partial c^2}{\partial y_i}+\chi_j\frac{\partial^2 c^0}{\partial x_i\partial x_j}
  +\frac{\partial \bar c^1}{\partial x_i})  \right]\ n_i = 0
    \hspace{0,5cm}\hbox{over $\Gamma$.}\label{CD_BCdefc2}& 
\end{numcases}
We observe that the solution must depend on three forcing terms, which are associated with ${\partial^2 c^0}/{\partial x_j \partial x_k} $,
${\partial c^0}/{\partial x_j}$
and ${\partial \bar c^1}/{\partial x_j} $, respectively. By virtue of linearity, the solution is a linear combination of particular solutions associated with each
of the three forcing terms. Note that the problem linked to ${\partial \bar c^1}/{\partial x_j} $ is identical to that observed at the first order for
${\partial c^0}/{\partial x_j}$ in the boundary value problem which defines $c^1$  (Eqs. (\ref{CD_Eq5_2bis})-(\ref{CD_Eq7_2bis})). Therefore,
the solution reads
\begin{equation}
 c^2= \eta_{jk}(\overrightarrow{y})\frac{\partial^2 c^0}{\partial x_j \partial x_k}
 +\pi_j(\overrightarrow{y})\frac{\partial c^0}{\partial x_j}
 +\chi_j(\overrightarrow{y}) \frac{\partial \bar c^1}{\partial x_j}
 +\bar c^2(\overrightarrow{x}, t),
 \label{CD_defc2}
\end{equation}
where $\bar c^2(\overrightarrow{x}, t)$ is an arbitrary function and where
\begin{numcases}{}
 <\eta_{jk}>=0,
 \\
 <\pi_j>=0.
\end{numcases}
The detailed definitions of $\eta_{jk}$ and $\pi_j$ are reported in Appendix \ref{def_eta_pi}.
%
\subsubsection{Derivation of the first corrector}
%
At the fourth order, the boundary-value problem made of Eqs. (\ref{CD_Eq2_adim})-(\ref{CD_Eq4_adim})  yields:
\begin{numcases}{}
 \frac{\partial c^1}{\partial t}
 -\frac{\partial}{\partial y_i}\left[ D_0 (\frac{\partial c^3}{\partial y_i}+\frac{\partial c^2}{\partial x_i})
 - c^0v_i^2 -c^1v_i^1-c^2v_i^0\right]&\nonumber\\
 -\frac{\partial}{\partial x_i}\left[D_0 (\frac{\partial c^2}{\partial y_i}+\frac{\partial c^1}{\partial x_i})-c^0v_i^1-c^1v_i^0  \right]=0
 \hspace{0,5cm}\hbox{within $\Omega_\mathrm{p} $,}\label{CD_Eq5_4}&\\
 v_i^0=v_i^1=v_i^2=0
 \hspace{0,5cm}\hbox{over $\Gamma$,}\\
 \left[ D_0 (\frac{\partial c^3}{\partial y_i}+\frac{\partial c^2}{\partial x_i})\right]n_i=0
 \hspace{0,5cm}\hbox{over $\Gamma$.}\label{CD_Eq7_4}
\end{numcases}
The first corrector of the macroscopic description is obtained by integrating
 Eq. (\ref{CD_Eq5_4}) over $\Omega_\mathrm p$. This leads to
 \begin{equation}
  \phi\frac{\partial \bar c^1}{\partial t}-\frac{\partial}{\partial x_i}<D_0 (\frac{\partial c^2}{\partial y_i}+\frac{\partial c^1}{\partial x_i}) >
  +\frac{\partial}{\partial x_i}<c^0v_i^1+
  c^1 v_i^0>=0.
   \label{2nd_order__corr_macro_00}
 \end{equation}
 Using the expressions obtained for $c^1$ and $c^2$, Eqs. (\ref{D_defc1}) and  (\ref{CD_defc2}), we get
\begin{equation}
 \frac{\partial c^2}{\partial y_i}+\frac{\partial c^1}{\partial x_i}=
 \gamma_{ijk}^1\frac{\partial^2c^0}{\partial x_j\partial x_k}+\frac{\partial \pi_j}{\partial y_i}\frac{\partial c^0}{\partial x_j}
 +\gamma_{ij}^0\frac{\partial \bar c^1}{\partial x_j},
 \label{gradyc2_plus_gradxc1}
\end{equation}
with
\begin{equation}
 \gamma_{ijk}^1=\frac{\partial \eta_{jk}}{\partial y_i}+\chi_i \delta_{jk}.
 \label{def_gamma1}
\end{equation} 
Then, noticing that 
\begin{equation}
 \frac{\partial}{\partial x_i}<c^0v_i^1+
  c^1 v_i^0> = <v_i^1 >\frac{\partial c^0}{\partial x_i}+\frac{\partial}{\partial x_i}\left[<v_i^0 \chi_j  >\frac{\partial c^0}{\partial x_j}   \right]
  + <v_i^0 >\frac{\partial \bar c^1}{\partial x_i},
\end{equation}
Eq. (\ref{2nd_order__corr_macro_00}) becomes:
\begin{equation}
 \phi\frac{\partial\bar c^1}{\partial t}
 -\frac{\partial}{\partial x_i}(E_{ijk}\frac{\partial^2 c^0}{\partial x_j\partial x_k}+
 D'_{ij}\frac{\partial c^0}{\partial x_j}
 +D_{ij}\frac{\partial \bar c^1}{\partial x_j})
 +<v_i^1>\frac{\partial c^0}{\partial x_i}+ <v_i^0>\frac{\partial \bar c^1}{\partial x_i}
 =0,
 \label{2nd_order__corr_macro_01}
\end{equation}
where
\begin{numcases}{}
 E_{ijk}= <D_0 (\frac{\partial \eta_{jk}}{\partial y_i}+\chi_i \delta_{jk}) >= <D_0 \gamma^1_{ijk}>,
 \label{def_E}\\
 D'_{ij}= <D_0\frac{\partial \pi_j}{\partial y_i}-v_i^0 \chi_j>.
 \label{def_D'}
\end{numcases}
The third-order tensor $E_{ijk}$ is symmetric with respect to its last two indices and antisymmetric with respect to its first two indices
({\em Cf.} Appendix \ref{Prop_E}). Note further that $E_{ijk}$ can be determined from vector $\chi_i$, without determining tensor $\eta_{jk}$ ({\em Cf.} Appendix \ref{Prop_E}). As a result of the  antisymmetry property of $E_{ijk}$, the second-order gradient term of
Eq. (\ref{2nd_order__corr_macro_01}) vanishes. Thus, the first corrector finally reads:
\begin{equation}
 \phi\frac{\partial\bar c^1}{\partial t}
 -\frac{\partial}{\partial x_i}(
 D'_{ij}\frac{\partial c^0}{\partial x_j}
 +D_{ij}\frac{\partial \bar c^1}{\partial x_j})
 +<v_i^1>\frac{\partial c^0}{\partial x_i}+ <v_i^0>\frac{\partial \bar c^1}{\partial x_i}
 =0.
 \label{2nd_order__corr_macro}
\end{equation}
From its definition Eq. (\ref{def_D'}), we see that the second-order tensor $D'_{ij}$ contains a convective term: it is therefore a dispersion tensor.
It is a non-symmetric tensor 
which can be decomposed into a symmetric and  an antisymmetric parts ({\em Cf.} Appendix  \ref{Prop_D'}). Furthermore, it can be determined from vectors $v_i^0$ and $\chi_j$, without solving boundary value problem Eqs. (\ref{CD_Eq5_4})-(\ref{CD_Eq7_4}) ({\em Cf.} Appendix  \ref{Prop_D'}).
\subsubsection{Second-order macroscopic description}
Let add Eq. (\ref{CD_1st_macro_1}) to Eq. (\ref{2nd_order__corr_macro}) multiplied by $\varepsilon$. We get:
\begin{numcases}{}
  \phi\displaystyle\frac{\partial}{\partial t}(c^0 + \varepsilon \bar c^1)
  -\displaystyle\frac{\partial}{\partial x_i}\left[D_{ij}\displaystyle\frac{\partial}{\partial x_j} (c^0 + \varepsilon \bar c^1)
  +\varepsilon D'_{ij}\displaystyle\frac{\partial c^0}{\partial x_j}\right]
  \vspace{0,2cm}\nonumber\\
  +(<v_i^0> + \varepsilon <v_i^1>)\displaystyle\frac{\partial c^0}{\partial x_i}+\varepsilon <v_i^0>\displaystyle\frac{\partial \bar c^1}{\partial x_i}=0.
 \label{2ndord_macro_des_0}
\end{numcases}
Defining the second-order macroscopic concentration and average fluid velocity by
\begin{numcases}{}
 <c> = <c^0> + \varepsilon\ {\bar c}^1 +{\mathcal O}(\varepsilon^2 < c>),
&\\
 <\overrightarrow{v}>=<\overrightarrow{v}^0>+\varepsilon <\overrightarrow {v}^1>+{\mathcal O}(\varepsilon^2 <\overrightarrow {v}>),&
\end{numcases}
the second-order macroscopic description is written as follows
\begin{numcases}{}
 \phi\displaystyle\frac{\partial <c>}{\partial t}-\displaystyle\frac{\partial}{\partial x_i} \left[(D_{ij}+\varepsilon D'_{ij})
 \displaystyle\frac{\partial <c>}{\partial x_j}  \right]
 + <v_i>\displaystyle\frac{\partial <c>}{\partial x_i}
 \vspace{0,2cm}\nonumber\\
 = {\mathcal O} (\varepsilon^2 \phi \displaystyle\frac{\partial <c>}{\partial t}).
 \label{2nd_ord_model_adim}
\end{numcases}
When cast in dimensional variables, Eq. (\ref{2nd_ord_model_adim}) becomes
\begin{numcases}{}
 \phi\displaystyle\frac{\partial <\hat c>}{\partial \hat t}
 - \displaystyle\frac{\partial}{\partial X_i} \left[(\hat D_{ij}^{\hbox{\tiny diff}}+ \hat {D'}_{ij}^{\hbox{\tiny eff}})\displaystyle\frac{\partial <\hat c>}{\partial X_j}  \right]
 +\ <\hat v_i>\displaystyle\frac{\partial <\hat c>}{\partial X_i}
 \vspace{0,2cm}\nonumber\\
 = {\mathcal O} (\varepsilon^2 \phi \displaystyle\frac{\partial <\hat c>}{\partial \hat t}),
 \label{2nd_ord_model_dim}
\end{numcases}
where
\begin{equation}
 \hat {D'}_{ij}^{\mathrm{eff}} = D_\mathrm c\varepsilon D'_{ij}.
\end{equation}
The second-order fluid velocity is such that ({\em Cf.} Section \ref{hom_fluid_flow}):
\begin{numcases}{}
 <\hat v_i >= - \displaystyle\frac{\hat N_{ijk}^{\hbox{\tiny eff}}}{\hat\mu}\displaystyle\frac{\partial^2 <\hat p>}{\partial \hat X_j\partial \hat X_k}
 - \displaystyle\frac{\hat K_{ij}^{\hbox{\tiny eff}}}{\hat\mu}\displaystyle\frac{\partial <\hat p>}{\partial \hat X_j}
 + {\mathcal O}(\varepsilon^2 <\hat v_i >),&\label{Darcy_2nd_order}\\
 \displaystyle\frac{\partial <\hat v_i >}{\partial \hat X_i}={\mathcal O}(\varepsilon^2 \displaystyle\frac{\partial <\hat v_i >}{\partial \hat X_i}).&
\end{numcases}
Note that combining both above equations leads to:
\begin{equation}
 \displaystyle\frac{\partial}{\partial \hat X_i}(\displaystyle\frac{\hat K_{ij}^{\hbox{\tiny eff}}}{\hat\mu}\displaystyle\frac{\partial <\hat p>}{\partial \hat X_j})
 ={\mathcal O}(\varepsilon^2 \displaystyle\frac{\partial <\hat v_i >}{\partial \hat X_i}).
\end{equation}
Therefore, the second-order macroscopic transport description is a model of advection-dispersion, in which the tensor of dispersion 
is non-symmetric ({\em Cf.} Appendix  \ref{Prop_D'}) and follows  property Eq. (\ref{pheno-tensor_disp}) of the phenomenological model of dispersion.
The fluid velocity verifies a second-order law Eq. (\ref{Darcy_2nd_order}), 
which reduces to Darcy's law in case of an isotropic medium. In other words, the second-order macroscopic transport model is similar to the
phenomenological dispersion transport equation Eq. (\ref{pheno-model_disp}). 
%
\subsection{Third-order homogenisation}
%
\label{Transp_3rdorder_hom}
\subsubsection{Boundary value problem for  $c^3$}
%
The fourth-order boundary value problem, Eqs. (\ref{CD_Eq5_4})-(\ref{CD_Eq7_4}), can be transformed into the following boundary value problem
for $c^3$ (Cf. Appendix \ref{BVP_c3}):
\begin{numcases}{}
  \displaystyle\frac{\partial}{\partial y_i}\left[D_0(\displaystyle\frac{\partial c^3}{\partial y_i}+\eta_{jk}\displaystyle\frac{\partial^3 c^0}{\partial x_i \partial x_j \partial x_k}
  +\pi_j \displaystyle\frac{\partial^2 c^0}{\partial x_i \partial x_j}+\chi_j \displaystyle\frac{\partial^2 \bar c^1}{\partial x_i \partial x_j}
  +\displaystyle\frac{\partial \bar c^2}{\partial x_i})  \right]=\vspace{0,2cm}\nonumber\\
   (\displaystyle\frac{1}{\phi}\chi_i D_{jk}-D_0 \gamma_{ijk}^1)\displaystyle\frac{\partial^3c^ 0}{\partial x_i \partial x_j \partial x_k}\vspace{0,2cm}\nonumber\\
 + (v_i^0 \gamma_{ijk}^1- D_0 \displaystyle\frac{\partial \pi_k}{\partial y_j}+ \displaystyle\frac{1}{\phi} D'_{jk}-\displaystyle\frac{1}{\phi}\chi_j <v_k^0>)
 \displaystyle\frac{\partial^2 c^0}{\partial x_j \partial x_k}\vspace{0,2cm}\nonumber\\
 +(\displaystyle\frac{1}{\phi}D_{ij}-D_0\gamma_{ij}^0) \displaystyle\frac{\partial^2 \bar c^1}{\partial x_i \partial x_j}\vspace{0,2cm}\nonumber\\
 + (v_i^0\displaystyle\frac{\partial \pi_j}{\partial y_i}+v_i^1 \gamma_{ij}^0- \displaystyle\frac{1}{\phi}\chi_i \displaystyle\frac{\partial <v_j^0>}{\partial x_i}-\displaystyle\frac{1}{\phi} <v_j^1>)
 \displaystyle\frac{\partial c^0}{\partial x_j}\vspace{0,2cm}\nonumber\\
 +(v_i^0\gamma_{ij}^0-\displaystyle\frac{1}{\phi}<v_j^0>)\displaystyle\frac{\partial \bar c^1}{\partial x_j}
 \hspace{0,5cm}\hbox{in $\Omega_{\mathrm p} $,}
 \label{defc3_eq1}\\
%
 \left[D_0(\displaystyle\frac{\partial c^3}{\partial y_i}+\eta_{jk}\displaystyle\frac{\partial^3 c^0}{\partial x_i \partial x_j \partial x_k}
  +\pi_j \displaystyle\frac{\partial^2 c^0}{\partial x_i \partial x_j}+\chi_j \displaystyle\frac{\partial^2 \bar c^1}{\partial x_i \partial x_j}
  +\displaystyle\frac{\partial \bar c^2}{\partial x_i})  \right]n_i = 0\vspace{0,2cm}\nonumber\\
   \hspace{0,5cm}\hbox{on $\Gamma$.}
   \label{defc3_eq2}
\end{numcases}
From the  above boundary value problem and its variational formulation ({\em Cf.} Appendix \ref{c3_var_form} ), it can be seen that the solution must 
depend on the following forcing terms:
${\partial^3 c^0}/{\partial x_j \partial x_k \partial x_l}$, ${\partial^2 c^0}/{\partial x_k\partial x_l} $,
${\partial^2 \bar c^1}/{\partial x_k \partial x_l}$, ${\partial c^0}/{\partial x_j}$, ${\partial \bar c^1}/{\partial x_j}$
and $ {\partial \bar c^2}/{\partial x_j}$. We note that the problem linked to $ {\partial \bar c^2}/{\partial x_j}$ is identical to that associated with 
${\partial c^0}/{\partial x_j}$ in the boundary value problem for $c^1$ Eqs. (\ref{CD_Eq5_2bis})-(\ref{CD_Eq7_2bis}). Furthermore, 
the problem associated with ${\partial \bar c^1}/{\partial x_j}$
is identical to that linked to ${\partial c^0}/{\partial x_j}$  in the boundary value problem for $c^2$, Eqs. (\ref{CD_eqdefc2})-(\ref{CD_BCdefc2}),
and the problem linked to ${\partial^2 \bar c^1}/{\partial x_k \partial x_l}$
is identical to that obtained for ${\partial^2 c^0}/{\partial x_k\partial x_l} $ in the boundary value problem for $c^2$.
Consequently, the solution reads:
\begin{equation}
\begin{array}{l}
 c^3 = \xi_{jkl}(\overrightarrow{y})\displaystyle\frac{\partial^3 c^0}{\partial x_j \partial x_k \partial x_l}+
 \tau_{kl}(\overrightarrow{y})\displaystyle\frac{\partial^2 c^0}{\partial x_k\partial x_l}
 +\eta_{kl}(\overrightarrow{y})\displaystyle\frac{\partial^2 \bar c^1}{\partial x_k \partial x_l}\vspace{0,2cm}\\
 + \theta_j(\overrightarrow{y}) \displaystyle\frac{\partial c^0}{\partial x_j}
 +\pi_j(\overrightarrow{y}) \displaystyle\frac{\partial \bar c^1}{\partial x_j}
 +\chi_j (\overrightarrow{y})\displaystyle\frac{\partial \bar c^2}{\partial x_j}+\bar c^3 (\overrightarrow{x}, t),
\end{array}
\label{CD_defc3}
\end{equation}
where $\bar c^3 (\overrightarrow{x}, t)$ is an arbitrary function, and where
\begin{numcases}{}
 < \xi_{jkl} >=0,\\
 <\tau_{kl}>=0,\\
 <\theta_j >=0.
\end{numcases}
The exact definitions of $\xi_{jkl}$, $\tau_{kl} $ and $\theta_j$ are reported in Appendices \ref{def_xi}, \ref{def_tau} and \ref{def_theta}, respectively. 
Let us recall that $\chi_j$ is related to the definition of $c^1$ Eq. (\ref{D_defc1}), while $\eta_{jk}$ and $\pi_j $ have been introduced
in the definition of $c^2$ Eq. (\ref{CD_defc2}). Note that in expression Eq. (\ref{CD_defc3}), $\xi_{jkl}$, $\eta_{jk}$, $\chi_j$ 
are only related to the diffusion mechanism, while
$\tau_{kl}$, $\theta_j$ and $\pi_j$ contain  both  diffusive and  convective terms.
\subsubsection{Derivation of the second corrector}
Let now consider
the boundary-value problem Eqs. (\ref{CD_Eq2_adim})-(\ref{CD_Eq4_adim})  at the fifth order:
\begin{numcases}{}
 \displaystyle\frac{\partial c^2}{\partial t}
 -\displaystyle\frac{\partial}{\partial y_i}\left[ D_0 (\displaystyle\frac{\partial c^4}{\partial y_i}+\displaystyle\frac{\partial c^3}{\partial x_i})
 - c^0v_i^3 -c^1v_i^2-c^2v_i^1-c^3v_i^0\right]&\nonumber\\
 -\displaystyle\frac{\partial}{\partial x_i}\left[D_0 (\displaystyle\frac{\partial c^3}{\partial y_i}+\displaystyle\frac{\partial c^2}{\partial x_i})-c^0v_i^2-c^1v_i^1-c^2v_i^0  \right]=0
 \hspace{0,5cm}\hbox{within $\Omega_\mathrm{p} $,}\label{CD_Eq5_5}&\\
 \left[ D_0 (\displaystyle\frac{\partial c^4}{\partial y_i}+\displaystyle\frac{\partial c^3}{\partial x_i})\right]n_i=0
 \hspace{0,5cm}\hbox{over $\Gamma$.}\label{CD_Eq7_5}
\end{numcases}
Integrating Eq. (\ref{CD_Eq5_5}) over $\Omega_\mathrm p$, we get: 
\begin{equation}
  \phi\displaystyle\frac{\partial \bar c^2}{\partial t}-\displaystyle\frac{\partial}{\partial x_i}<D_0 (\displaystyle\frac{\partial c^3}{\partial y_i}+\displaystyle\frac{\partial c^2}{\partial x_i}) >
  +\displaystyle\frac{\partial}{\partial x_i}<c^0v_i^2+
  c^1 v_i^1+c^2v_i^0>=0.
   \label{3rd_order__corr_macro_00}
 \end{equation}
 Using Eqs. (\ref{CD_defc2}) and (\ref{CD_defc3}), we deduce that
\begin{numcases}{}
   \displaystyle\frac{\partial c^3}{\partial y_i}+\displaystyle\frac{\partial c^2}{\partial x_i}=
    \gamma_{ijkl}^2\displaystyle\frac{\partial^3 c^0}{\partial x_j\partial x_k \partial x_l}\vspace{0,2cm}\nonumber\\
     +(\displaystyle\frac{\partial \tau_{jk}}{\partial y_i}+\pi_i \delta_{jk})\displaystyle\frac{\partial^2 c^0}{\partial x_j\partial x_k}
     +\gamma_{ijk}^1\displaystyle\frac{\partial^2\bar c^1}{\partial x_j\partial x_k}\vspace{0,2cm} \label{CD_gradyc3_plus_gradxc2_1}\\
      +\displaystyle\frac{\partial \theta_j}{\partial y_i}\displaystyle\frac{\partial c^0}{\partial x_j}
  +\displaystyle\frac{\partial\pi_j}{\partial y_i}\displaystyle\frac{\partial \bar c^1}{\partial x_j}
  +\gamma_{ij}^0\displaystyle\frac{\partial\bar c^2}{\partial x_j},\nonumber
\end{numcases}
where
\begin{equation}
 \gamma_{ijkl}^2=\displaystyle\frac{\partial\xi_{jkl}}{\partial y_i}+\eta_{ij}\delta{kl}.
\end{equation}
Then, noticing that:
\begin{numcases}{}
 \displaystyle\frac{\partial}{\partial x_i}<c^0v_i^2+
  c^1 v_i^1+c^2v_i^0>=\vspace{0,2cm}\nonumber\\
  \displaystyle\frac{\partial}{\partial x_i}\left[<v_i^0 \eta_{jk} >\displaystyle\frac{\partial^2 c^0}{\partial x_j\partial x_k}
  +<v_i^1\chi_j+v_i^0\pi_j>\displaystyle\frac{\partial c^0}{\partial x_j}+<v_i^0\chi_j >\displaystyle\frac{\partial \bar c^1}{\partial x_j}  \right]\vspace{0,2cm}\\
  + <v_i^2 >\displaystyle\frac{\partial c^0}{\partial x_i}+<v_i^1 >\displaystyle\frac{\partial \bar c^1}{\partial x_i}+<v_i^0 >\displaystyle\frac{\partial \bar c^2}{\partial x_i},\nonumber
\end{numcases}
Eq. (\ref{3rd_order__corr_macro_00}) becomes:
\begin{numcases}{}
  \phi \displaystyle\frac{\partial \bar c^2}{\partial t} -\displaystyle\frac{\partial}{\partial x_i}[
  F_{ijkl}\displaystyle\frac{\partial^3 c^0}{\partial x_j\partial x_k \partial x_l}
  +E'_{ijk} \displaystyle\frac{\partial^2 c^0}{\partial x_j\partial x_k}
  +E_{ijk}\displaystyle\frac{\partial^2 \bar c^1}{\partial x_j\partial x_k}\vspace{0,2cm}\nonumber\\
  +D''_{ij}\displaystyle\frac{\partial c^0}{\partial x_j}
  +D'_{ij}\displaystyle\frac{\partial \bar c^1}{\partial x_j}
  +D_{ij}\displaystyle\frac{\partial \bar c^2}{\partial x_j}]\vspace{0,2cm}\label{CD_3rd_macro_1}\\
  +<v_i^2>\displaystyle\frac{\partial c^0}{\partial x_i}+<v_i^1>\displaystyle\frac{\partial \bar c^1}{\partial x_i} + <v_i^0>\displaystyle\frac{\partial \bar c^2}{\partial x_i}
  =0,\nonumber
\end{numcases}
where 
\begin{numcases}{}
 F_{ijkl}= <D_0 \displaystyle\frac{\partial \xi_{jkl}}{\partial y_i}+\eta_{ij}\delta_{kl}>,&\label{def_F}\\
 E'_{ijk}=<D_0\displaystyle\frac{\partial \tau_{jk}}{\partial y_i}-v_i^0\eta_{jk} >,&\label{def_E_prime}\\
 D''_{ij}=<D_0\displaystyle\frac{\partial \theta_j}{\partial y_i}-v_i^1\chi_j -v_i^0\pi_j   >.\label{def_D_dbleprime}
\end{numcases}
Tensor
$F_{ijkl}$ is a fourth-order tensor of diffusion. It can be calculated from vector $\overrightarrow{\chi}$ and tensor $\bar{\bar\eta} $,
without solving the boundary-value problem Eqs. (\ref{defc3_eq1})-(\ref{defc3_eq2})
({\em Cf.} Appendix \ref{Prop_F}). The third-order tensor $E'_{ijk}$ and the second-order tensor $D''_{ij}$ are tensors of dispersion. They can also be determined without solving the boundary-value problem Eqs. (\ref{defc3_eq1})-(\ref{defc3_eq2}) ({\em Cf.} Appendices \ref{Prop_D''} and \ref{Prop_E'}). Finally, we conclude that the second corrector can be determined from $\bar{\bar\eta} $, $\overrightarrow{\chi}$, $\overrightarrow{\pi}$, $\overrightarrow{v}^0$ and $\overrightarrow{v}^1$.
\subsubsection{Third-order macroscopic description}
Let add Eq. (\ref{2ndord_macro_des_0}) to Eq. (\ref{CD_3rd_macro_1}) multiplied by  $\varepsilon^2$:
\begin{numcases}{}
  \phi\displaystyle\frac{\partial}{\partial t}(c^0 + \varepsilon\bar c^1 +\varepsilon^2\bar c^2)\vspace{0,2cm}\nonumber\\
  -\displaystyle\frac{\partial}{\partial x_i}[D_{ij}\displaystyle\frac{\partial }{\partial x_j}(c^0 + \varepsilon \bar c^1 +\varepsilon^2\bar c^2)
  + \varepsilon D'_{ij}\displaystyle\frac{\partial}{\partial x_j}(c^0 + \varepsilon \bar c^1)+ \varepsilon^2 D''_{ij}\displaystyle\frac{\partial c^0}{\partial x_j}
  \vspace{0,2cm}\nonumber\\
  +\varepsilon^2 E'_{ijk}\displaystyle\frac{\partial^2 c^0}{\partial x_j\partial x_k}+\varepsilon^2 F_{ijkl}\displaystyle\frac{\partial^3 c^0}{\partial x_j\partial x_k\partial x_l}
  ]\vspace{0,2cm} \label{2ndord_macrodesc}\\
  + (<v_i^0> +\varepsilon <v_i^1> +\varepsilon^2 <v_i^2>)\displaystyle\frac{\partial c^0}{\partial x_i}\vspace{0,2cm}\nonumber\\
  + \varepsilon (<v_i^0> +\varepsilon <v_i^1>) \displaystyle\frac{\partial \bar c^1}{\partial x_i}+\varepsilon^2 <v_i^0> \displaystyle\frac{\partial \bar c^2}{\partial x_i}=0.
  \nonumber
\end{numcases}
Defining the third-order macroscopic concentration and fluid velocity by
\begin{numcases}{}
 <c> = <c^0> + \varepsilon {\bar c}^1 +\varepsilon^2 \bar c^2+{\mathcal O}(\varepsilon^3 < c>),
 &\\
 <\vec {v}>=<\vec { v}^0>+\varepsilon <\vec {v}^1>+\varepsilon^2 <\vec {v}^2>+{\mathcal O}(\varepsilon^3 <\vec {v}>),
 &
\end{numcases}
the third-order macroscopic description is written as follows
\begin{numcases}{}
 \phi\displaystyle\frac{\partial <c>}{\partial t}-\displaystyle\frac{\partial}{\partial x_i} \left[(D_{ij}+\varepsilon D'_{ij}+\varepsilon^2 D''_{ij})
 \displaystyle\frac{\partial <c>}{\partial x_j}  \right]
 \vspace{0,2cm}\nonumber\\
 -\displaystyle\frac{\partial}{\partial x_i} \left[ \varepsilon^2 E'_{ijk}\displaystyle\frac{\partial^2 <c>}{\partial x_j\partial x_k}
 +\varepsilon^2 F_{ijkl}\displaystyle\frac{\partial^3 <c>}{\partial x_j\partial x_k\partial x_l}\right]\vspace{0,2cm}\label{3rdd_ord_model_adim}\\
 + <v_i>\displaystyle\frac{\partial <c>}{\partial x_i}= {\mathcal O} (\varepsilon^3 \phi \displaystyle\frac{\partial <c>}{\partial t}).\nonumber
\end{numcases}
In dimensional variables, we get:
\begin{numcases}{}
 \phi\displaystyle\frac{\partial <\hat c>}{ \partial \hat t}-
 \displaystyle\frac{\partial}{\partial X_i} \left[(\hat D^{\hbox{\tiny diff}}_{ij}
 +\hat D^{'\hbox{\tiny disp}}_{ij}
 +\hat D^{''\hbox{\tiny disp}}_{ij})
 \displaystyle\frac{\partial <\hat c>}{\partial X_j}  \right]
 \vspace{0,2cm}\nonumber\\
 -\displaystyle\frac{\partial}{\partial X_i} \left[\hat E^{'\hbox{\tiny disp}}_{ijk}\displaystyle\frac{\partial^2 <\hat c>}{\partial X_j\partial X_k}
 + \hat F^{\hbox{\tiny diff}}_{ijkl}\displaystyle\frac{\partial^3 <\hat c>}{\partial X_j\partial X_k\partial X_l}\right]\vspace{0,2cm}\label{3rdd_ord_model_dim}\\
 + <\hat v_i>\displaystyle\frac{\partial <\hat c>}{\partial X_i}= {\mathcal O} (\varepsilon^3 \phi \displaystyle\frac{\partial <\hat c>}{\partial \hat t}),\nonumber
\end{numcases}
where
\begin{numcases}{}
 \hat D^{''\hbox{\tiny disp}}_{ij}=D_\mathrm c\varepsilon^2 D''_{ij},
 &\\
 \hat E^{'\hbox{\tiny disp}}_{ijk}=\varepsilon l D_\mathrm c E'_{ijk},
 &\\
 \hat F^{\hbox{\tiny diff}}_{ijkl}= l^2 D_\mathrm c F_{ijkl}.
\end{numcases}
The third-order fluid velocity verifies ({\em Cf.} Section \ref{hom_fluid_flow}):
\begin{numcases}{}
 <\hat v_i >= - \displaystyle\frac{\hat P_{ijkl}^{\hbox{\tiny eff}}}{\hat\mu}\displaystyle\frac{\partial^2 <\hat p>}{\partial \hat X_j\partial \hat X_k\partial \hat X_l}
 - \displaystyle\frac{\hat N_{ijk}^{\hbox{\tiny eff}}}{\hat\mu}\displaystyle\frac{\partial^2 <\hat p>}{\partial \hat X_j\partial \hat X_k}
 - \displaystyle\frac{\hat K_{ij}^{\hbox{\tiny eff}}}{\hat\mu}\displaystyle\frac{\partial <\hat p>}{\partial \hat X_j}
 + {\mathcal O}(\varepsilon^3 <\hat v_i >),&\label{Darcy_3rd_order}\\
 \displaystyle\frac{\partial <\hat v_i >}{\partial \hat X_i}={\mathcal O}(\varepsilon^3 \displaystyle\frac{\partial <\hat v_i >}{\partial \hat X_i}).&
\end{numcases}
Note that when combining both above equations, the second-gradient term vanishes, due the antisymmetry property of tensor  $\hat N_{ijk}^{\hbox{\tiny eff}}$.\\
\\The third-order transport model Eq. (\ref{3rdd_ord_model_dim}) introduces a fourth-order tensor of diffusion, and a third-order and an additional second-order tensors of dispersion.
%
\section{Macroscopic fluxes}
%
\label{macro_fluxes}
\setcounter{equation}{0}
\subsection{Volume vs surface averages}
With the homogenisation  averaging procedure, macroscopic descriptions are expressed in terms of  variables which are systematically  defined as volume averages.
Specifying the meaning of the macroscopic variables, {\em i.e.} determining whether the use of volume averages is appropriate or not
is thus  an important issue \citep{Hass79,Cos05,Hill72}. 
In the particular context of solute transport in porous media, since a solute flux is physically defined over a specific area,
macroscopic fluxes should thus be defined as surface averages. 
\subsection{Writing of local and homogenised equations in terms of fluxes }
In order to address the above described issue,  we may rewrite the local and the homogenised equations in terms of fluxes.
We shall thus rewrite Eq. (\ref{CD_Eq2_adim})  as follows
\begin{equation}
 \displaystyle\frac{\partial c}{\partial t}+\overrightarrow{\nabla}\cdot \overrightarrow{q} = 0
  \hspace{0,5cm}\hbox{within $\Omega_\mathrm p $},
  \label{CD_Eq2_adim_flux}
\end{equation}
where the local flux $\overrightarrow{q} $ is defined by
\begin{equation}
 \overrightarrow{q} = - D_0\overrightarrow{\nabla}  c+  c \overrightarrow{v}.
 \label{CD_Eq1_adim_flux}
\end{equation}
The no-flux boundary condition now reads
\begin{equation}
 \overrightarrow{q} \cdot \overrightarrow{n}=0
 \hspace{0,5cm}\hbox{over $\Gamma$}.
  \label{CD_nofluxbc_flux}
\end{equation}
Flux $ \overrightarrow{q}$ is looked for in the form of the following asymptotic expansion in powers of $\varepsilon$
\begin{equation}
 \overrightarrow{q}=\overrightarrow{q}^0 (\overrightarrow{y}, \overrightarrow{x}) + \varepsilon \overrightarrow{q}^1 (\overrightarrow{y}, \overrightarrow{x})
 + \varepsilon^2 \overrightarrow{q}^2 (\overrightarrow{y}, \overrightarrow{x}) + ...
\end{equation}
This leads to the following perturbations equations for Eqs. (\ref{CD_Eq2_adim_flux})-(\ref{CD_Eq1_adim_flux}) at the successive orders of powers of $\varepsilon$:
\begin{numcases}{}
 \overrightarrow{q}^0= - D_0 (\overrightarrow{\nabla}_y  c^1 + \overrightarrow{\nabla}_x  c^0 ) + c^0 \overrightarrow{v}^0&\label{Flux_eq_1_1}\\
 \overrightarrow{q}^1=  - D_0 (\overrightarrow{\nabla}_y  c^2 + \overrightarrow{\nabla}_x  c^1 ) + c^0 \overrightarrow{v}^1  
 + c^1 \overrightarrow{v}^0   &\label{Flux_eq_1_2}\\
 \overrightarrow{q}^2= - D_0 (\overrightarrow{\nabla}_y  c^3 + \overrightarrow{\nabla}_x  c^2 ) + c^0 \overrightarrow{v}^2
 + c^1 \overrightarrow{v}^1 + c^2 \overrightarrow{v}^0\label{Flux_eq_1_3}
\end{numcases}
and 
\begin{numcases}{}
 \overrightarrow{\nabla}_y\cdot \overrightarrow{q}^0=0&\label{Flux_eq_2_1}\\
 \displaystyle\frac{\partial c^0}{\partial t}+  \overrightarrow{\nabla}_y\cdot \overrightarrow{q}^1+
  \overrightarrow{\nabla}_x\cdot \overrightarrow{q}^0=0&\label{Flux_eq_2_2}\\  
  \displaystyle\frac{\partial c^1}{\partial t}+  \overrightarrow{\nabla}_y\cdot \overrightarrow{q}^2+
  \overrightarrow{\nabla}_x\cdot \overrightarrow{q}^1=0&\label{Flux_eq_2_3}
\end{numcases}
As for the homogenised equations at the first three orders, Eqs. (\ref{D_1st_ord_macro_0}), (\ref{2nd_order__corr_macro}) and (\ref{CD_3rd_macro_1}),
they are re-expressed as follows\\
\\\underline{First-order}
\begin{numcases}{}
 \phi \displaystyle\frac{\partial c^0}{\partial t} + \displaystyle\frac{\partial <q_i^0 >} {\partial x_i} = 0&\label{flux_1st_ord_macro_massbal}\\
 <q_i^0 > = - D_{ij} \displaystyle\frac{\partial c^0}{\partial x_j} + c^0 <v_i^0>\label{flux_1st_ord_macro_moy_q1}
\end{numcases}
\underline{Second-order corrector}
\begin{numcases}{}
  \phi \displaystyle\frac{\partial \bar c^1}{\partial t} + \displaystyle\frac{\partial <q_i^1 >} {\partial x_i} = 0&\label{flux_2nd_ord_macro_massbal}\\
  <q_i^1 >= - E_{ijk} \displaystyle\frac{\partial^2 c^0}{\partial x_j \partial x_k} -D'_{ij}\displaystyle\frac{\partial c^0}{\partial x_j}
  - D_{ij} \displaystyle\frac{\partial \bar c^1}{\partial x_j}
  + c^0 <v_i^1 > + \bar c^1 <v_i^0 >\label{flux_2nd_ord_macro_moy_q1}
\end{numcases}
\underline{Third-order corrector}
\begin{numcases}{}
 \phi \displaystyle\frac{\partial \bar c^2}{\partial t} + \displaystyle\frac{\partial <q_i^2 >} {\partial x_i} = 0&\\
  <q_i^2 > = - F_{ijkl} \displaystyle\frac{\partial^3 c^0}{\partial x_j \partial x_k \partial x_l}
  - E'_{ijk} \displaystyle\frac{\partial^2 c^0}{\partial x_j \partial x_k}
  - E_{ijk} \displaystyle\frac{\partial^2 \bar c^1}{\partial x_j \partial x_k}&\nonumber\\
  - D''_{ij} \displaystyle\frac{\partial c^0}{\partial x_j} - D'_{ij} \displaystyle\frac{\partial \bar c^1}{\partial x_j}
  - D_{ij } \displaystyle\frac{\partial \bar c^2}{\partial x_j} + c^0 <v_i^2 >+  c^1 <v_i^1 > + c^2 <v_i^0 > \label{flux_2nd_ord_macro_moy_q2}  
\end{numcases}
To analyse whether volume averages  of  local  fluxes have the properties of macroscopic fluxes,
we consider the following identity to transform volume averages into surface averages \citep{Aur05}
\begin{equation}
 \displaystyle\frac{\partial}{\partial y_i}(y_j q_i)\equiv y_j\displaystyle\frac{\partial q_i}{\partial y_i}+q_j.
 \label{identity}
\end{equation}
%
\subsection{First-order macroscopic flux}
%
\label{1srt_ord_macro_flux}
Let take $ q_i = q_i^0$ 
in Eq. (\ref{identity}) and then integrate over $\Omega_p$. 
Since by Eq. (\ref{Flux_eq_2_1}) $q_i^0$ is solenoidal  according to $\overrightarrow{y}$, it reduces to
\begin{equation}
 \displaystyle\frac{1}{\mid\Omega\mid}\int_{\Omega_\mathrm p} \displaystyle\frac{\partial}{\partial y_i}(y_j q_i^0)\ d\Omega =<q_j^0 >.
\end{equation}
Applying the divergence theorem and the no-flux boundary condition Eq. (\ref{CD_nofluxbc_flux}) of order $\varepsilon^0$, leads to:
\begin{equation}
\displaystyle\frac{1}{\mid\Omega\mid}\int_{\delta\Omega_p\cap\delta\Omega}\ y_j q_i^0 n_i\ dS = <q_j^0 >.
\end{equation}
\begin{figure}[ht!!]
    \begin{center}
    \includegraphics[width=7cm]{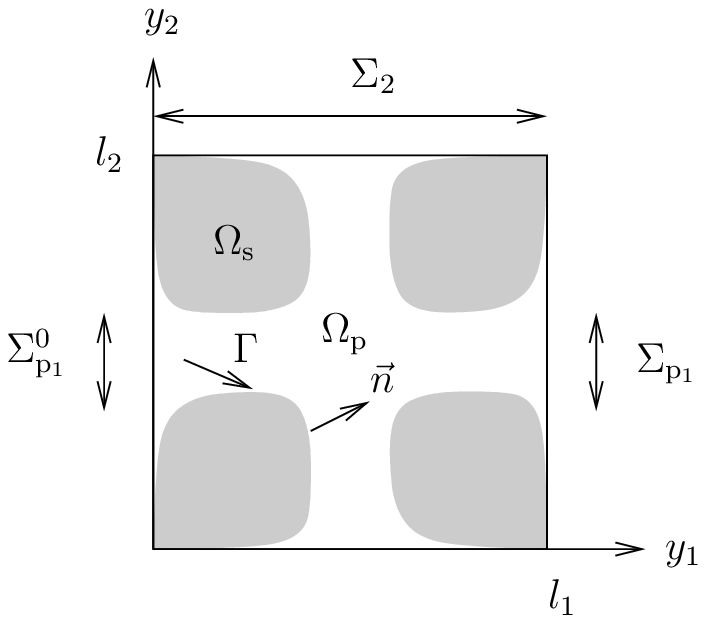}
    \end{center}
  \caption{\label{fig_medium_fluxes}{\it\small Two-dimensional periodic cell  $\Omega$.}}
  \end{figure}
Let $l_i$ be the dimensionless length of the period along the $y_i$ axis. We denote by
$\Sigma_i^0$ and  
$\Sigma_i$  the cross-sections of the period at $y=0$ and  $y_i=l_i e_i$, respectively.
$\Sigma_{p_i}^0$ and $\Sigma_{p_i}$ are the fluid parts of $\Sigma_i^0$ and $\Sigma_i$, respectively
({\em Cf.} Fig.\ref{fig_medium_fluxes}).
We firstly note that $y_j q_i^0$ is $\Omega$-periodic in the $y_k (k\neq j)$ direction. Consequently, only integrals over boundaries
$\Sigma_j^0$ and  
$\Sigma_j$ (where the normal unit vectors are $\pm e_j $) remain, the others cancel out. Furthermore, $y_j q_i^0=0$ for $y_j=0$. 
Therefore, the integral over $\Sigma_j^0$ is zero. We are left with
\begin{equation}
 \displaystyle\frac{1}{\mid\Omega\mid}\int_{\delta\Omega_p\cap\delta\Omega}\ y_j q_i^0 n_i\ dS=
 \displaystyle\frac{1}{\mid\Omega\mid}\int_{\Sigma_{p_j}} l_j q_i^0\ dS= \displaystyle\frac{1}{\mid\Sigma_j\mid}\int_{\Sigma_{p_j}}  q_j^0\ dS,
\end{equation}
(without summation over $j$), and we define
\begin{equation}
 <q_j^0 >_{\Sigma_i}= \displaystyle\frac{1}{\mid\Sigma_j\mid}\int_{\Sigma_{p_j}}  q_j^0\ dS.
\end{equation}
Hence, we have 
\begin{equation}
 <q_j^0> = <q_j^0>_{\Sigma_j},
\end{equation}
which means that the volume average of $q_j^0$ is equal to a surface average. Therefore, $<q_j^0>$ has the properties of a  macroscopic flux.
As a consequence, from the expression of $\overrightarrow{q}^0$, Eq. (\ref{Flux_eq_1_1}), we deduce that
\begin{equation}
 < v_j^0> = <v_j^0>_{\Sigma_j},
 \label{v0_volmean_surf_mean}
\end{equation}
which means that the volume average of $\overrightarrow{v}^0$ has the properties of a Darcy's velocity. 
Note that the equalities between  volume averages and  surface averages of $q_j^0$ and $v_j^0$ 
are consequences of the solenoidal character of $\overrightarrow{q}^0$ 
and $\overrightarrow{v}^0$,
according to variable $\overrightarrow{y}$.\\
Therefore, Eqs. (\ref{flux_1st_ord_macro_massbal})-(\ref{flux_1st_ord_macro_moy_q1}) can be rewritten as
\begin{numcases}{}
 \phi\displaystyle\frac{\partial c^0}{\partial t} + \displaystyle\frac{\partial <q_i^0 >_{\Sigma_{p_i}}}{\partial x_i}=0,\label{macro_flux_1st_ord_macro_massbal}&\\
 <q_i^0 >_{\Sigma_{p_i}}= - D_{ij} \displaystyle\frac{\partial c^0}{\partial x_j} + c^0 <v_i^0 >_{\Sigma_{p_i}},\label{macro_flux_1st_ord_macro_moy_q1}
\end{numcases}
and the first-order macroscopic description Eq. (\ref{CD_1st_macro_adim}) can be expressed as
\begin{numcases}{}
 \phi \displaystyle\frac{\partial <c>}{\partial t} + \displaystyle\frac{\partial <q_i >_{\Sigma_{p_i}}}{\partial x_i}
 ={\mathcal O}(\varepsilon \phi \displaystyle\frac{\partial <c>}{\partial t}),&\label{CD_1st_macro__flux_adim_eq1}\\
  <q_i >_{\Sigma_{p_i}}= - D_{ij} \displaystyle\frac{\partial <c>}{\partial x_j} + <c> <v_i >_{\Sigma_{p_i}}
  +{\mathcal O}(\varepsilon <q_i >_{\Sigma_{p_i}})\label{CD_1st_macro__flux_adim_eq2},
\end{numcases}
where the first-order macroscopic solute flux and fluid velocity are defined by
\begin{numcases}{}
 <q_i >_{\Sigma_{p_i}}= <q_i^0 >_{\Sigma_{p_i}} + {\mathcal O}(\varepsilon <q_i >_{\Sigma_{p_i}}),&\\
 <v_i>_{\Sigma_{p_i}} = <v_i^0 >_{\Sigma_{p_i}} + {\mathcal O}(\varepsilon <v_i >_{\Sigma_{p_i}}).&
\end{numcases}
Finally, in dimensional variables the first-order transport model read
\begin{numcases}{}
 \phi \displaystyle\frac{\partial <\hat c>}{\partial \hat t} + \displaystyle\frac{\partial <\hat q_i >_{\hat \Sigma_{p_i}}}{\partial \hat X_i}
 ={\mathcal O}(\varepsilon \phi \displaystyle\frac{\partial <\hat c>}{\partial \hat t}),&\label{CD_1st_macro__flux_dim_eq1}\\
  <\hat q_i >_{\hat\Sigma_{p_i}}= - \hat D^{\hbox{\tiny diff}}_{ij} \displaystyle\frac{\partial <\hat c>}{\partial \hat X_j} + <\hat c> <\hat v_i>_{\hat \Sigma_{p_i}}
  +{\mathcal O}(\varepsilon <\hat q_i >_{\hat \Sigma_{p_i}})\label{CD_1st_macro__flux_dim_eq2}.
\end{numcases}
\subsection{Second-order macroscopic flux}
%
\label{Macro_fluxes_2nd_ord}
To analyse the volume average of $\overrightarrow{q}^1$, let consider identity Eq. (\ref{identity}) with $q_i=q_i^1$
and integrate over $\Omega_\mathrm p$. This yields
\begin{equation}
 <q_i^1>_{\Sigma_{\mathrm p_i}}=<y_i\displaystyle\frac{\partial q_j^1}{\partial y_j}>+<q_i^1>.
 \label{def_surf_mean_qi1}
\end{equation}
Now, by Eq. (\ref{Flux_eq_2_2}), we get that $\overrightarrow{q}^1$ is non-solenoidal
\begin{equation}
 \displaystyle\frac{\partial q_j^1}{\partial y_j}=-\displaystyle\frac{\partial q_j^0}{\partial x_j}-\displaystyle\frac{\partial c^0}{\partial t}.
 \label{partialq1partialy_1}
\end{equation}
Consequently, the volume average of $\overrightarrow{q}^1$ is not equal to its surface average
\begin{equation}
 <q_i^1>_{\Sigma_{\mathrm p_i}}\neq <q_i^1>,
\end{equation}
which means that $<\overrightarrow{q}^1>$ is not a macroscopic flux.\\
By starting from Eq. (\ref{def_surf_mean_qi1}) and then using Eq. (\ref{partialq1partialy_1}) to get the term $<y_i{\partial q_j^1}/{\partial y_j}> $,
we obtain the following expression for $<\overrightarrow{q}^1>_{\Sigma_{p_i}}$
(Cf. Appendix \ref{Det_surf_av_q1}):
\begin{numcases}{}
 <q_i^1>_{\Sigma_{\mathrm p_i}}=
  - (E_{ijk}-E_{ijk}^{\Sigma})\displaystyle\frac{\partial^2 c^0}{\partial x_j\partial x_k}\vspace{0,2cm}\nonumber\\
  -(D'_{ij}-{D'}_{ij}^{\Sigma})\displaystyle\frac{\partial c^0}{\partial x_j}- D_{ij}\displaystyle\frac{\partial \bar c^1}{\partial x_j}\vspace{0,2cm}\label{q1surface_mean_final}\\
  +c^0<v_i^1>_{\Sigma_{\mathrm p_i}} + \bar c^1 <v_i^0>_{\Sigma_{\mathrm p_i}},\nonumber
\end{numcases}
where
\begin{numcases}{}
 E_{ijk}^{\Sigma}=<D_0 y_i\gamma_{jk}^0 - \displaystyle\frac{1}{\phi}y_iD_{jk} >,
 \label{def_ESigma}
\\
 {D'}_{ij}^{\Sigma} = <y_i(\displaystyle\frac{1}{\phi}<v_j^0>-v_j^0  )>.
 \label{def_D'Sigma}
\end{numcases}
Using Eq. (\ref{def_surf_mean_qi1}),  the first corrector of the macroscopic description, Eq. (\ref{flux_2nd_ord_macro_massbal}),
can be rewritten  in terms of the second-order macroscopic flux  as follows:
\begin{equation}
 \phi\displaystyle\frac{\partial \bar c^1}{\partial t}+\displaystyle\frac{\partial}{\partial x_i} ( <q_i^1 >_{\Sigma_{p_i}})=
 \displaystyle\frac{\partial}{\partial x_i}(<y_i \displaystyle\frac{\partial q_j^1}{\partial y_j}>).
\end{equation}
Then, using Eqs. (\ref{av_yipartialq1ipartialyj}), (\ref{av_v1Sigma_vs_av_v1Omega}), (\ref{def_ESigma}), (\ref{def_D'Sigma}), it becomes
\begin{numcases}{}
 \phi\displaystyle\frac{\partial \bar c^1}{\partial t}+\displaystyle\frac{\partial}{\partial x_i} ( <q_i^1 >_{\Sigma_{p_i}})=\vspace{0,2cm}\nonumber\\
 \displaystyle\frac{\partial}{\partial x_i}[E_{ijk}^{\Sigma}\displaystyle\frac{\partial^2 c^0}{\partial x_j \partial x_k}+ D_{ij}^{'\Sigma}\displaystyle\frac{\partial c^0}{\partial x_j}
 - c^0 (<v_i^1> - <v_i^1>_{\Sigma_{p_i}})].
 \label{2ndord_correc_avSigmaq1}
\end{numcases}
Now, in order to obtain the corresponding second-order macroscopic description, let firstly  add Eq. (\ref{macro_flux_1st_ord_macro_massbal}) 
to Eq. (\ref{2ndord_correc_avSigmaq1}) multiplied by $\varepsilon$. We get
\begin{numcases}{}
\phi\displaystyle\frac{\partial <c>}{\partial t}+\displaystyle\frac{\partial <q_i>_{\Sigma_{p_i}}}{\partial x_i}=\vspace{0,2cm}\nonumber\\
\displaystyle\frac{\partial}{\partial x_i}\left[\varepsilon E_{ijk}^{\Sigma}\displaystyle\frac{\partial^2 <c>}{\partial x_j\partial x_k}
+\varepsilon D_{ij}^{'\Sigma}\displaystyle\frac{\partial <c>}{\partial x_j}- <c>(<v_i>-<v_i>_{\Sigma_{p_i}})\right]\vspace{0,2cm}\label{2ndord_macrodesc_avqSigma}\\
+{\mathcal O} (\varepsilon^2 \phi\displaystyle\frac{\partial <c>}{\partial t}).\nonumber
\end{numcases}
Next, we add Eq. (\ref{macro_flux_1st_ord_macro_moy_q1}) to Eq. (\ref{q1surface_mean_final}) multiplied $\varepsilon$, and we obtain
\begin{numcases}{}
  <q_i>_{\Sigma_{p_i}}=- \varepsilon (E_{ijk}-E_{ijk}^{\Sigma})\displaystyle\frac{\partial^2 <c>}{\partial x_j x_k}\vspace{0,2cm}\nonumber\\
   -(D_{ij}+\varepsilon D'_{ij}-\varepsilon D_{ij}^{'\Sigma})\displaystyle\frac{\partial <c>}{\partial x_j}\vspace{0,2cm}\label{2ndord_macro_flux}\\
   + <c> <v_i>_{\Sigma_{p_i}} + {\mathcal O}(\varepsilon^2 <q_i>_{\Sigma_{p_i}}).\nonumber
\end{numcases}
In the above equations, the second-order macroscopic solute flux and fluid velocity are defined by
\begin{numcases}{}
 <q_i>_{\Sigma_{p_i}} = <q_i^0>_{\Sigma_{p_i}} + \varepsilon <q_i^1>_{\Sigma_{p_i}} +{\mathcal O}(\varepsilon^2 <q_i>_{\Sigma_{p_i}}),
 &\\
<v_i>_{\Sigma_{p_i}} = <v_i^0>_{\Sigma_{p_i}} + \varepsilon <v_i^1>_{\Sigma_{p_i}} +{\mathcal O}(\varepsilon^2 <v_i>_{\Sigma_{p_i}}),
 &
\end{numcases}
respectively. In dimensional variables, Eqs. (\ref{2ndord_macrodesc_avqSigma}) and (\ref{2ndord_macro_flux}) read
\begin{numcases}{}
  \phi\displaystyle\frac{\partial <\hat c>}{\partial \hat t}
  + \displaystyle\frac{\partial <\hat q_i>_{{\hat\Sigma}_{p_i}}}{\partial \hat X_i}=\vspace{0,2cm}\nonumber\\
  \displaystyle\frac{\partial}{\partial \hat X_i}\left[
  \hat E_{ijk}^{\Sigma}\displaystyle\frac{\partial^2 <\hat c>}{\hat \partial X_j\partial X_k} 
  + \hat D_{ij}^{'\Sigma}\displaystyle\frac{\partial <\hat c>}{\partial \hat X_j}
  - <\hat c>(<\hat v_i>-<\hat v_i>_{\hat\Sigma_{p_i}} \right]\vspace{0,2cm}\nonumber\\
  + {\mathcal O} (\varepsilon^2 \phi\displaystyle\frac{\partial <\hat c>}{\partial \hat t}),\vspace{0,5cm}\label{2nd_ord_macro_mass_bal_surf_flux}\\
  <\hat q_i>_{\Sigma_{p_i}}=- (\hat E^{\hbox{\tiny diff}}_{ijk}-\hat E_{ijk}^{\Sigma})\displaystyle\frac{\partial^2 <\hat c>}{\partial X_j X_k}
  -(\hat D^{\hbox{\tiny diff}}_{ij}+\hat D^{'\hbox{\tiny disp}}_{ij}-\hat D_{ij}^{'\Sigma})\displaystyle\frac{\partial <\hat c>}{\partial X_j}\vspace{0,2cm}\nonumber\\
  +<\hat c> <\hat v_i>_{\Sigma_{p_i}} + {\mathcal O}(\varepsilon^2 <\hat q_i>_{\Sigma_{p_i}}),\label{2nd_ord_macro_surf_flux}
\end{numcases}
where 
\begin{numcases}{}
{\hat E}^{\hbox{\tiny diff}}_{ijk}=l D_c E_{ijk},\\
 {\hat E}_{ijk}^{\Sigma}=l D_c E_{ijk}^{\Sigma},\\
 {\hat D}_{ij}^{'\Sigma}=\varepsilon D_c D_{ij}^{'\Sigma}.
\end{numcases}
%
\subsection{Third-order macroscopic flux}
%
Proceeding in the same manner as in \S\ref{Macro_fluxes_2nd_ord}, we also conclude
that 
\begin{equation}
 <\overrightarrow{q}^2 > \neq  <\overrightarrow{q}^2 >_{\Sigma_{p_i}},
\end{equation}
and we show that in dimensional variables, the third-order transport model expressed in terms of the macroscopic flux reads (Cf. Appendix \ref{Det_surf_av_q2}):
\begin{numcases}{}
 \phi \displaystyle\frac{\partial <\hat c >}{\partial \hat t}+\displaystyle\frac{\partial}{\partial X_i}(<q_i >_{\Sigma_{p_i}})=\vspace{0,2cm}\nonumber\\
 \displaystyle\frac{\partial}{\partial X_i}\left[
 \hat F^{\Sigma}_{ijkl}\displaystyle\frac{\partial^3 <c>}{\partial X_j \partial X_k\partial X_l}
 + (\hat E^{\Sigma}_{ijk}+\hat E^{'\Sigma}_{ijk})\displaystyle\frac{\partial^2 <c>}{\partial X_j\partial X_k}
 \right.\vspace{0,2cm}\nonumber\\
 \left. + (\hat D^{'\Sigma}_{ij}+\hat D^{''\Sigma}_{ij})\displaystyle\frac{\partial <c>}{\partial X_j} + <c> (<v_i>-<v_i>_{\Sigma_{p_i}})\right]\vspace{0,2cm}\nonumber\\
 + {\mathcal O}(\varepsilon^3 \phi \displaystyle\frac{\partial <\hat c >}{\partial \hat t}).\vspace{0,5cm}\label{3rd_ord_macro_mass_bal_surf_flux}\\
  <\hat q_i>_{\Sigma_{p_i}}=- (\hat F_{ijkl}^{\hbox{\tiny diff}}-\hat F^{\Sigma}_{ijk})\displaystyle\frac{\partial^3 <c>}{\partial X_j\partial X_k\partial X_l}\vspace{0,2cm}\nonumber\\
  -(\hat E_{ijk}^{\hbox{\tiny diff}}-\hat E^{\Sigma}_{ijk}+\hat E'_{ijk}-\hat E^{'\Sigma}_{ijk})\displaystyle\frac{\partial^2 <c>}{\partial X_j\partial X_k}\vspace{0,2cm}\nonumber\\
  - (\hat D_{ij}+\hat D'_{ij}-\hat D^{'\Sigma}_{ij}+\hat D^{''}_{ij}-\hat D^{''\Sigma}_{ij})\displaystyle\frac{\partial <c>}{\partial X_j}\vspace{0,2cm}\nonumber\\
  + <\hat c> <\hat v_i>_{\Sigma_{p_i}}+{\mathcal O}(\varepsilon^3 <\hat q_i >_{\Sigma_{p_i}})\label{3rd_ord_macro_surf_flux}
\end{numcases}
where (Cf. Appendix \ref{Det_surf_av_q2})
\begin{numcases}{}
\hat F^{\Sigma}_{ijkl}= l^2 D_c F^{\Sigma}_{ijkl},
\\
\hat E^{'\Sigma}_{ijk}= \varepsilon l D_c E^{'\Sigma}_{ijk},
\\
\hat D^{''\Sigma}_{ij}=\varepsilon^2 D_c D^{''\Sigma}_{ij}.
\end{numcases}
Since $<\hat q_i>_{\Sigma_{p_i}} $ has the properties of a macroscopic flux, the  right-hand-sides of  the mass-balance equations,  Eq. (\ref{2nd_ord_macro_mass_bal_surf_flux}) 
and Eq. (\ref{3rd_ord_macro_mass_bal_surf_flux}), represent source terms,
which are actually expressions of the second-order and third-order non-local effects, respectively. 
\section{Conclusions}
\label{conclu}
\setcounter{equation}{0}
In the present paper, higher-order asymptotic homogenisation up to the third order of solute transport in the advective-diffusive regime is performed.
The main result of the study is that low scale separation induces dispersion
effects. At the second order, the transport model is similar to the classical model of dispersion: the dispersion
tensor is the sum of the diffusion tensor and a mechanical dispersion tensor, while this property is not verified in the
homogenised dispersion model obtained at higher P\'eclet number. The velocity is governed by a second-order law which reduces to Darcy’s law in case of isotropy.
Thus, the second-order model of advection-diffusion is similar to the phenomenological model of dispersion.
The third-order description contains second and third concentration gradient terms, 
with a fourth order tensor of diffusion and with a third-order and an additional second-order tensors of dispersion. Hence, these results show that when employing the first order model while $\varepsilon$ is  not ``very'' small would, for example,  lead to a wrong estimate of the tensor of effective diffusion from experimental data. We generally admit  that a first-order model, whose degree of  precision is ${\mathcal O}(\varepsilon)$, is valid for a value of $\varepsilon$ up to $\varepsilon \approx 0.1$. Consequently, we may estimate that the p-order model is required when $\varepsilon^p\approx 0.1 $. 
The analysis of the macroscopic  fluxes shows that the second and the third order macroscopic fluxes are distinct from the volume averages of the corresponding local fluxes.
From the writing of the second and third order models in terms of the macroscopic fluxes arise expressions of the non-local effects.
All theses results are valid for macroscopically homogeneous media and  macroscopic
heterogeneity would lead to stronger non-local effects.\\
The results at Péclet number  ${\mathcal O}(\varepsilon)$ can quite easily be deduced from the above analysis. This leads to the model of diffusion at the first order,
the model of advection-diffusion
at the second order and dispersion effects appear at the third order.
Eventually, we may conclude  that scale separation is a crucial issue whenever the fluid is in motion, since
low scale separation induces a modification of the apparent transport regime \citep{Roy18}.\\
An important property of higher-order homogenised models is that edge effects are induced:
the boundary layer created by the heterogeneity  may
affect the
homogenised solution inside the domain  in higher orders with respect to $\varepsilon$. 
Numerical simulations of the  above derived effective higher-order equations thus requires a specific treatment of these edge effects \citep{Smy00,Bua01,Dum90}.
A discussion on that topic is complex and beyond the scope of this paper.\\
Since the advection-diffusion equation is a Fokker-Planck type equation,
the higher-order transport homogenised equations may appear to be similar to a generalised Fokker-Planck equation \citep{Ris89}.
Such equation, which describes the time evolution of a probability density function  is obtained by a Kramers–Moyal expansion which transforms an 
integro differential master equation.
\cite{Paw67} has proved
that finite truncations of the generalised Fokker-Planck equation at
any order greater than the second leads to a logical inconsistency, as the function must then have 
negative values at least for sufficiently small times and in isolated regions.
This argument may be used to put into question the validity of higher-order homogenised transport models \citep{Mau91}.
In this regard the work of  \cite{vanKam81} provides the
framework for the introduction of a small parameter which allows for the construction of a modified Kramers-Moyal expansion.
Then, one can approximate the expansion by a finite number of terms which involves derivatives of order higher than two, using an appropriate perturbation technique. In this case, the contribution from higher-order terms diminishes, because of their
order in the small parameter. 
Such an expansion  is admittedly questionable in view of Pawula’s theorem, but can be controlled when manipulated with care \citep{Popes15}. Thus, the theorem of Pawula does not necessarily restrict the truncation of higher order terms, when we can formally obtain high-order perturbative
equations \citep{Kan17} and nonvanishing higher-order coefficients have been observed in various systems \citep{Anv16, Fri11, Pru07, Tut04, Kim08, Pet09, Pet15}.
Therefore, though higher-order perturbative models might, in some cases, have negative values at some isolated times and positions, this does not invalidate the models derived in the study, which are valid only in zones where large concentration gradients are applied.

%
\appendix
%
\section*{Appendices}
\section{First-order homogenisation}
\setcounter{equation}{0}
\subsection{Definition of vector ${\chi_j}$}
\label{1st_ord_localprob}
Let multiply the local problem defined by  Eqs. (\ref{CD_Eq5_2bis})-(\ref{CD_Eq7_2bis}) by a test function $\alpha$
satisfying the condition of having zero average, 
and then, let integrate over $\Omega_\mathrm p$.
We obtain the following variational formulation
\begin{equation}
  \int_{\Omega_\mathrm p}\ D_0\displaystyle\frac{\partial \alpha}{\partial y_i}\displaystyle\frac{\partial c^1}{\partial y_i}\ d\Omega = 
  - \int_{\Omega_\mathrm p}\ D_0\displaystyle\frac{\partial \alpha}{\partial y_i}\ d\Omega\ \displaystyle\frac{\partial c^0}{\partial x_i}.
  \label{D_look_weakform_c1_2}
\end{equation}
Vector $\chi_j$ is the solution for $c^1$ when ${\partial c^0}/{\partial x_i}=\delta_{ij}$.
Therefore, the variational formulation associated with $\chi_j$
is
\begin{equation}
 \int_{\Omega_\mathrm p}\ D_0\displaystyle\frac{\partial \alpha}{\partial y_i}\displaystyle\frac{\partial \chi_j}{\partial y_i}\ d\Omega = 
  - \int_{\Omega_\mathrm p}\ D_0 \displaystyle\frac{\partial \alpha}{\partial y_j}\ d\Omega,
  \label{form_var_chi}
\end{equation}
and $\chi_j$ must satisfy
\begin{equation}
 \left\lbrace
 \begin{array}{l}
 \displaystyle\displaystyle\frac{\partial}{\partial y_i}\left[D_0(\displaystyle\frac{\partial \chi_j}{\partial y_i}+ \delta_{ij})  \right]=0
 \hspace{0.5cm}\hbox{in $\Omega_\mathrm{p} $,}\vspace{0.2cm}\\
 \left[D_0(\displaystyle\displaystyle\frac{\partial \chi_j}{\partial y_i}+\delta_{ij})  \right]\ n_i=0\hspace{0.5cm}\hbox{on $\Gamma$,}\vspace{0,2cm}\\
 <\chi_j> = 0,\vspace{0,2cm}\\
\hbox{$ \overrightarrow{\chi}$: periodic in $\overrightarrow{y}$.}
 \end{array}
\right.
\label{D_def_chi}
\end{equation}
%
\subsection{Symmetry of tensor $D_{ij}$}
%
To demonstrate the symmetry of $D_{ij}$, we firstly take
\label{prop_D}
\begin{equation}
 \left\lbrace
 \begin{array}{l}
 \alpha = \chi_q,\\
 c^1=\chi_p, \displaystyle\displaystyle\frac{\partial c^0}{\partial x_i}=\delta_{ip},
 \end{array}
 \right.
\end{equation}
into Eq. (\ref{D_look_weakform_c1_2}). This leads to:
\begin{equation}
\displaystyle\int_{\Omega_\mathrm p}\ D_0\displaystyle\frac{\partial \chi_q}{\partial y_i}\displaystyle\frac{\partial \chi_p}{\partial y_i}\ d\Omega = 
-\displaystyle\int_{\Omega_\mathrm p}\ D_0\displaystyle\frac{\partial \chi_q}{\partial y_p}\ d\Omega.
\label{Sym_D_Eq1}
 \end{equation}
Next, we consider
\begin{equation}
 \left\lbrace
 \begin{array}{l}
 \alpha = \chi_p,\\
 c^1=\chi_q, \displaystyle\frac{\partial c^0}{\partial x_i}=\delta_{iq},
 \end{array}
 \right.
\end{equation}
into Eq. (\ref{D_look_weakform_c1_2}), which leads to:
\begin{equation}
\displaystyle\int_{\Omega_\mathrm p}\ D_0\displaystyle\frac{\partial \chi_p}{\partial y_i}\displaystyle\frac{\partial \chi_q}{\partial y_i}\ d\Omega = 
-\displaystyle\int_{\Omega_\mathrm p}\ D_0\displaystyle\frac{\partial \chi_p}{\partial y_q}\ d\Omega.
\label{Sym_D_Eq2}
 \end{equation}
 By Eqs. (\ref{Sym_D_Eq1}) and (\ref{Sym_D_Eq2}), we deduce that:
 \begin{equation}
  \displaystyle\int_{\Omega_\mathrm p}\ \displaystyle\frac{\partial \chi_q}{\partial y_p}\ d\Omega=
  \displaystyle\int_{\Omega_\mathrm p}\ \displaystyle\frac{\partial \chi_p}{\partial y_q}\ d\Omega.
 \end{equation}
Consequently, we have:
\begin{equation}
 D_{qp}=D_{pq},
\end{equation}
which proves the symmetry of  $\bar{\bar D}$.
%
\section{Second-order homogenisation}
%
\setcounter{equation}{0}
\subsection{Boundary value problem for $c^2$}
%
\label{BVP_c2}
The third-order boundary value given by Eqs. (\ref{CD_Eq5_3}) and (\ref{CD_Eq7_3}) can be written as follows:
\begin{numcases}{}
 \frac{\partial}{\partial y_i}\left[D_0 (\frac{\partial c^2}{\partial y_i}+\frac{\partial c^1}{\partial x_i})\right]
 -\frac{\partial}{\partial y_i} (c^0 v_i^1) -\frac{\partial}{\partial y_i}( c^1 v_i^0)=&\nonumber\\
  \frac{\partial c^0}{\partial t}-\frac{\partial}{\partial x_i}\left[D_0 (\frac{\partial c^1}{\partial y_i}+\frac{\partial c^0}{\partial x_i}) \right]
 +\frac{\partial}{\partial x_i}(c^0 v_i^0)
 \hspace{0,5cm}\hbox{within $\Omega_\mathrm{p}, $}\label{CD_ed_defc2_0}&\\
 \left[D_0(\frac{\partial c^2}{\partial y_i}+\frac{\partial c^1}{\partial x_i})  \right]n_i = 0
 \hspace{0,5cm}\hbox{over $\Gamma$}.
  \label{CD_Eq7_3bis}
\end{numcases}
Now, using Eq. (\ref{divyv0}), and the second order of  Eq. (\ref{F_eq2})
\begin{equation}
 \frac{\partial v_i^1}{\partial y_i}+\frac{\partial v_i^0}{\partial x_i}=0,
 \label{F_eq2_2}
\end{equation}
while bearing in mind Eq. (\ref{c0}),
the second and the third terms of the left hand side of Eq. (\ref{CD_ed_defc2_0}) can be transformed as follows:
\begin{numcases}{}
 \frac{\partial }{\partial y_i}(c^0 v_i^1)= - c^0\frac{\partial v_i^0}{\partial x_i},
 \label{gradyc0v1}\\
 \frac{\partial}{\partial y_i}(c^1 v_i^0)=
 v_i^0 \frac{\partial c^1}{\partial y_i}.
 \label{gradyc1v0}
\end{numcases}
Next, using  Eq. (\ref{CD_1st_macro_1}) we get
\begin{equation}
 \frac{\partial c^0}{\partial t}=
 \frac{1}{\phi}D_{ij} \frac{\partial^2 c^0}{\partial x_i\partial x_j}-\frac{1}{\phi}<v_i^0>\frac{\partial c^0}{\partial x_i},
 \label{CD_partialc0partialt}
\end{equation}
and from  Eq. (\ref{D_defc1}), we obtain
\begin{equation}
 \frac{\partial c^1}{\partial x_i}=\chi_j \frac{\partial^2c^0}{\partial x_i \partial x_j}+\frac{\partial \bar c^1}{\partial x_i}.
 \label{CD_gradxc1}
\end{equation}
Substituting Eqs. (\ref{gradyc0v1}) to (\ref{CD_gradxc1}) into  Eqs. (\ref{CD_ed_defc2_0})-(\ref{CD_Eq7_3bis}),
and then using the expression Eq. (\ref{gradyc1_plus_gradxc0}),
we get the boundary value problem Eqs. (\ref{CD_eqdefc2})-(\ref{CD_BCdefc2}).
%
\subsection{Definitions of tensor $\eta_{jk}$ and vector $\pi_j$}
%
\label{def_eta_pi}
By multiplying the local problem Eqs. (\ref{CD_eqdefc2})-(\ref{CD_BCdefc2}) 
by a test function $\alpha$ of zero average, and then integrating over $\Omega_\mathrm p$, we obtain its variational formulation: 
\begin{equation}
 \begin{array}{l}
 \displaystyle\int_{\Omega_\mathrm p}\  \frac{\partial \alpha}{\partial y_i} [D_0(\frac{\partial c^2}{\partial y_i}+ 
  \chi_j\ \frac{\partial^2 c^0}{\partial x_i \partial x_j}+ \frac{\partial \bar c^1}{\partial x_i})]\ d\Omega=\vspace{0,2cm}\\
   \displaystyle\int_{\Omega_\mathrm p}\  
  \alpha  D_0\ \gamma_{ij}^0 \ \frac{\partial^2 c^0}{\partial x_i \partial x_j}\ d\Omega
  - \displaystyle\int_{\Omega_\mathrm p}\  
  \alpha v_i^0\gamma_{ij}^0\frac{\partial c^0}{\partial x_j}\ d\Omega.
 \end{array}
 \label{var_form_2nd_ord_pb}
\end{equation}
$n_{lm}$ is the particular solution for $c^2$ when
\begin{numcases}{}
 \frac{\partial^2 c^0}{\partial x_i\partial x_j}=\delta_{il} \delta_{jm},&\nonumber\\
 \frac{\partial c^0}{\partial x_i}=\frac{\partial \bar c^1}{\partial x_i}=0.\nonumber
\end{numcases}
Therefore, the variational formulation associated with $n_{jk}$ reads
\begin{equation}
  \displaystyle\int_{\Omega_\mathrm p}\  \frac{\partial \alpha}{\partial y_i} D_0(\frac{\partial n_{lm}}{\partial y_i}+ 
  \chi_m\ \delta_{il})\ d\Omega=
   \displaystyle\int_{\Omega_\mathrm p}\  
  \alpha  D_0\ \gamma_{lm}^0 \ \ d\Omega,
 \label{form_var_eta}
\end{equation}
and $n_{jk}$ must satisfy
\begin{equation}
\left\lbrace
 \begin{array}{l}
  \displaystyle\frac{\partial}{\partial y_i}(D_0 (\displaystyle\frac{\partial\eta_{lm}}{\partial y_i}+\chi_m \delta_{il}))=
  \frac{1}{\phi}D_{lm}-D_0 \gamma_{lm}^0
  \hspace{0,5cm}\hbox{within $\Omega_\mathrm p$,}\vspace{0,2cm}\\
  (D_0 (\displaystyle\frac{\partial\eta_{lm}}{\partial y_i}+\chi_m \delta_{il}))n_i=0
   \hspace{0,5cm}\hbox{over $\Gamma$,}\vspace{0,2cm}\\
   <\eta_{lm}> = 0,\vspace{0,2cm}\\
   \hbox{$\eta_{lm}$: $\Omega$-periodic in variable $\overrightarrow{y}$}.
 \end{array}
\right.
\end{equation}
From its definition, we see that $\eta_{lm}$ is a parameter related to the diffusion mechanism.\\
\\$\pi_k$ is the solution for $c^2$ when
\begin{numcases}{}
 \frac{\partial c^0}{\partial x_j}=\delta_{jk},&\nonumber\\
 \frac{\partial \bar c^1}{\partial x_i}= \frac{\partial^2 c^0}{\partial x_i\partial x_j}=0.\nonumber
\end{numcases}
The variational formulation associated with $\pi_k$ is thus
\begin{equation}
 \displaystyle\int_{\Omega_\mathrm p}\  
 \frac{\partial \alpha}{\partial y_i} D_0 \frac{\partial \pi_k}{\partial y_i}\ d\Omega=
 - \displaystyle\int_{\Omega_\mathrm p}\  
  \alpha v_i^0 \gamma_{ik}^0 \ d\Omega,
\end{equation}
and $\pi_k$ must satisfy
\begin{equation}
 \left\lbrace
 \begin{array}{l}
  \displaystyle\frac{\partial}{\partial y_i}(D_0\frac{\partial\pi_k}{\partial y_i})=\gamma_{ik}^0v_i^0-\frac{1}{\phi}<v_k^0>
   \hspace{0,5cm}\hbox{within $\Omega_\mathrm p$,}\vspace{0,2cm}\\
    \displaystyle(D_0\frac{\partial\pi_k}{\partial y_i})n_i=0
     \hspace{0,5cm}\hbox{over $\Gamma$,}\vspace{0,2cm}\\
     <\pi_k>=0, \vspace{0,2cm}\\
      \hbox{$\pi_k$: $\Omega$-periodic in variable $\overrightarrow{y}$}.
 \end{array}
\right.
\end{equation}
From the above definition it is clear that vector $\overrightarrow{\pi}$ depends on both the diffusive and the convective phenomena, 
which characterises the presence of dispersive
effects.
\subsection{Properties of the third-order tensor $E_{ijk}$}
\label{Prop_E}
\subsubsection{Symmetry by construction of a third-order tensor with respect to its last two indices}
\label{3rdordTensors_sym_2_last_ind}
 By construction, $E_{ijk}$ is symmetric with respect its last two indices:
 \begin{equation}
  E_{ijk}\frac{\partial^2 c^0}{\partial x_j\partial x_k}=E_{ijk}\frac{\partial^2 c^0}{\partial x_k\partial x_j}=
  E_{ikj}\frac{\partial^2 c^0}{\partial x_k\partial x_j}=E_{ikj}\frac{\partial^2 c^0}{\partial x_j\partial x_k}.
 \end{equation}
Consequently:
\begin{equation}
  E_{ijk}= E_{ikj}.
  \label{sym_2_last_ind}
\end{equation}
In case of isotropy, third-order tensors are scalar multiples of the permutation tensor
\begin{equation}
 E_{ijk} = E\ \epsilon_{ijk}\hspace{0,5cm} \epsilon_{ijk}: \hbox{permutation tensor}.
\end{equation}
Since $\epsilon_{ijk}=-\epsilon_{ikj}$, Eq. (\ref{sym_2_last_ind}) induces that: $E=0$. Thus, any third-order tensor which is symmetric with respect to its last two indices is equal to zero in case of isotropy.

\subsubsection{Antisymmetry with respect to the first two indices}
Let take $\alpha=\eta_{lm}$ in the variational formulation associated with functions $\chi_j$ Eq. (\ref{form_var_chi}). We obtain
\begin{equation}
 \int_{\Omega_\mathrm p}\ D_0\frac{\partial \eta_{lm}}{\partial y_i}\frac{\partial \chi_j}{\partial y_i}\ d\Omega=
 -\int_{\Omega_\mathrm p}\ D_0\frac{\partial \eta_{lm}}{\partial y_j}\ d\Omega.
 \label{D_atisymE_eq1}
\end{equation}
Let now take $\alpha=\chi_j$ in the variational formulation associated with $\eta_{lm}$ Eq. (\ref{form_var_eta}). We get
\begin{equation}
\begin{array}{l}
 \displaystyle\int_{\Omega_\mathrm p}\ D_0\frac{\partial \eta_{lm}}{\partial y_i}\frac{\partial \chi_j}{\partial y_i}\ d\Omega=\vspace{0,2cm}\\
 -\displaystyle\int_{\Omega_\mathrm p}\ D_0\chi_m\frac{\partial\chi_j}{\partial y_l}\ d\Omega
 +\displaystyle\int_{\Omega_\mathrm p}\ D_0\chi_j\frac{\partial \chi_m}{\partial y_l}\ d\Omega.
 \end{array}
 \label{D_atisymE_eq2}
\end{equation}
From Eqs. (\ref{D_atisymE_eq1}) and (\ref{D_atisymE_eq2}), we deduce
\begin{equation}
  \displaystyle\int_{\Omega_\mathrm p}\ D_0(\frac{\partial \eta_{lm}}{\partial y_j}+\chi_j \delta_{lm})\ d\Omega=
  \displaystyle\int_{\Omega_\mathrm p}\ D_0\chi_m\frac{\partial\chi_j}{\partial y_l}\ d\Omega
  -\displaystyle\int_{\Omega_\mathrm p}\ D_0\chi_j\frac{\partial \chi_m}{\partial y_l}\ d\Omega.
\end{equation}
Thus, from the definition of $E_{jlm} $ Eq. (\ref{def_E}), we have
\begin{equation}
 E_{jlm}= \frac{1}{\mid\Omega\mid}\displaystyle\int_{\Omega_\mathrm p}\ D_0\chi_m\frac{\partial\chi_j}{\partial y_l}\ d\Omega
  -\frac{1}{\mid\Omega\mid}\displaystyle\int_{\Omega_\mathrm p}\ D_0\chi_j\frac{\partial \chi_m}{\partial y_l}\ d\Omega,
  \label{def_E_with_chi}
\end{equation}
and
\begin{equation}
 \begin{array}{l}
  E_{ljm}=\displaystyle \frac{1}{\mid\Omega\mid}\int_{\Omega_\mathrm p}\ D_0\chi_m\frac{\partial\chi_l}{\partial y_j}\ d\Omega
   - \frac{1}{\mid\Omega\mid}\displaystyle\int_{\Omega_\mathrm p}\ D_0\chi_l\frac{\partial \chi_m}{\partial y_j}\ d\Omega=\vspace{0,2cm}\\
  \displaystyle \frac{1}{\mid\Omega\mid} \displaystyle\int_{\Omega_\mathrm p}\ D_0\chi_j\frac{\partial \chi_m}{\partial y_l}\ d\Omega
   -\displaystyle \frac{1}{\mid\Omega\mid}\int_{\Omega_\mathrm p}\ D_0\chi_m\frac{\partial\chi_j}{\partial y_l}\ d\Omega.
 \end{array}
\end{equation}
Therefore
\begin{equation}
 E_{jlm}=-E_{ljm}.
\end{equation}
Since the medium is macroscopically homogeneous, $E_{ijk}$ does not depend on the macroscopic variable $\overrightarrow{x}$.
Consequently, the antisymmetry with respect to the two first indices implies that
\begin{equation}
 \frac{\partial }{\partial x_i}(E_{ijk}\frac{\partial^2 c^0}{\partial x_j\partial x_k})=0.
\end{equation}
From Eq. (\ref{def_E_with_chi}), we further note that tensor $E_{jlm}$ can be determined from vector $\overrightarrow\chi$.
\subsection{Properties of tensor $D'_{ij}$}
\label{Prop_D'}
Let take $c^2= \pi_k$ and $\alpha=\chi_l$ in the variational formulation of the second-order local problem Eq. (\ref{var_form_2nd_ord_pb}). We obtain
\begin{equation}
 \int_{\Omega_\mathrm p}\ \frac{\partial\chi_l}{\partial y_i}D_0\frac{\partial \pi_k}{\partial y_i}\ d \Omega = 
 -\int_{\Omega_\mathrm p}\chi_l\frac{\partial\chi_k}{\partial y_i}v_i^0\ d \Omega - \int_{\Omega_\mathrm p}\chi_l v_k^0\ d\Omega.
 \label{Prop_D'_eq1}
\end{equation}
Now, by taking $c^1=\chi_l$ and $\alpha=\pi_k$ in the variational formulation of the first-order problem Eq. (\ref{D_look_weakform_c1_2}), we get
\begin{equation}
 \int_{\Omega_\mathrm p}\frac{\partial \pi_k}{\partial y_i}D_0\frac{\partial\chi_l}{\partial y_i}\ d \Omega = 
 -\int_{\Omega_\mathrm p} D_0 \frac{\partial\pi_k}{\partial y_l}\ d\Omega.
  \label{Prop_D'_eq2}
\end{equation}
From Eqs. (\ref{Prop_D'_eq1}) and (\ref{Prop_D'_eq2}), we deduce
\begin{equation}
 \int_{\Omega_\mathrm p} D_0 \frac{\partial\pi_k}{\partial y_l}\ d\Omega=
 \int_{\Omega_\mathrm p}\chi_l\frac{\partial\chi_k}{\partial y_i}v_i^0\ d \Omega+
 \int_{\Omega_\mathrm p}\chi_l v_k^0\ d\Omega.
\end{equation}
Now, by considering the definition of $\bar{\bar D}'$, Eq. (\ref{def_D'}), with the above expression, it comes
\begin{equation}
 D'_{lk}= <D_0 \frac{\partial \pi_k}{\partial y_l}-v_l^0 \chi_k >=<\chi_l v_k^0>+<\chi_l\frac{\partial\chi_k}{\partial y_i}v_i^0> - <\chi_kv_l^0>,
 \label{D'_second_def}
\end{equation}
from which we deduce
\begin{equation}
 D'_{lk}-D'_{kl}=2 (<\chi_l v_k^0>-<\chi_k v_l^0>)\neq 0.
\end{equation}
Therefore, $\bar{\bar D}'$ is not symmetric:
\begin{equation}
 D'_{lk}\neq D'_{kl}.
\end{equation}
From Eq. (\ref{D'_second_def}), tensor $\bar{\bar D}'$ can be decomposed as
\begin{equation}
 D'_{lk}={^{s}D'}_{lk}+{^a D'}_{lk},
\end{equation}
where
\begin{equation}
 {^{s}D'}_{lk}=<\chi_l\frac{\partial\chi_k}{\partial y_i}v_i^0>
\end{equation}
is symmetric and where
\begin{equation}
 {^{a}D'}_{lk}=<\chi_l v_k^0>- <\chi_kv_l^0>
\end{equation}
is antisymmetric.\\
Furthermore, from Eq. (\ref{D'_second_def}), it can be seen that tensor $D'_{lk}$ can be determined from vectors $\overrightarrow \chi$ and $\overrightarrow v^0$.
%
\section{Third order homogenisation}
%
\setcounter{equation}{0}
\subsection{Boundary value problem for $c^3$}
%
\label{BVP_c3}
From  Eqs. (\ref{CD_Eq5_4})-(\ref{CD_Eq7_4}), we get the following boundary value problem for $c^3$:
\begin{numcases}{}
 \displaystyle\frac{\partial}{\partial y_i}\left[D_0 (\displaystyle\frac{\partial c^3}{\partial y_i}+\displaystyle\frac{\partial c^2}{\partial x_i})  \right]
 - \displaystyle\frac{\partial}{\partial y_i}(c^0 v_i^2)-\displaystyle\frac{\partial}{\partial y_i} (c^1v_i^1)-\displaystyle\frac{\partial}{\partial y_i} (c^2v_i^0)=
 \vspace{0,2cm}\nonumber\\
 \displaystyle\frac{\partial c^1}{\partial t}-\displaystyle\frac{\partial}{\partial x_i}
\left[ D_0 (\displaystyle\frac{\partial c^2}{\partial y_i}+\displaystyle\frac{\partial c^1}{\partial x_i})  \right]
  +\displaystyle\frac{\partial}{\partial x_i}(c^0 v_i^1)+\displaystyle\frac{\partial}{\partial x_i}(c^1 v_i^0)
 \hspace{0,2cm}\hbox{in $\Omega_\mathrm p, $}
 \label{CD_defc3_0}\\
 %
 \left[D_0 (\displaystyle\frac{\partial c^3}{\partial y_i}+\displaystyle\frac{\partial c^2}{\partial x_i}) 
 \right]n_i=0 \hspace{0,5cm}\hbox{over $\Gamma$.}
 \label{CD_bc_defc3}
\end{numcases}
Using Eq. (\ref{divyv0}) and Eq. (\ref{F_eq2_2}), and Eq. (\ref{F_eq2}) at the third order
\begin{equation}
 \displaystyle\frac{\partial v_i^2}{\partial y_i}+\displaystyle\frac{\partial v_i^1}{\partial x_i}=0,
\end{equation}
we deduce that
\begin{numcases}{}
 \displaystyle\frac{\partial}{\partial y_i}(c^0 v_i^2)=-c^0\displaystyle\frac{\partial v_i^1}{\partial x_i},
 \label{divyc0v2}\\
 \displaystyle\frac{\partial}{\partial y_i}(c^1v_i^1)=- c^1\displaystyle\frac{\partial v_i^0}{\partial x_i}+
 \displaystyle\frac{\partial c^1}{\partial y_i}v_i^1,
\\
 \displaystyle\frac{\partial}{\partial y_i}(c^2v_i^0)= \displaystyle\frac{\partial c^2}{\partial y_i}v_i^0,
 \\
 \displaystyle\frac{\partial}{\partial x_i}(c^0 v_i^1)= c^0 \displaystyle\frac{\partial v_i^1}{\partial x_i}+\displaystyle\frac{\partial c^0}{\partial x_i}v_i^1,
 \\
 \displaystyle\frac{\partial}{\partial x_i}(c^1 v_i^0)=c^1\displaystyle\frac{\partial v_i^0}{\partial x_i}+\displaystyle\frac{\partial c^1}{\partial x_i}v_i^0.
 \label{divxc1v0}
\end{numcases}
Then, substituting Eqs. (\ref{divyc0v2})-(\ref{divxc1v0}) into Eq. (\ref{CD_defc3_0}), 
while using Eqs. (\ref{gradyc1_plus_gradxc0}) and (\ref{gradyc2_plus_gradxc1}) yields
\begin{numcases}{}
   \displaystyle\frac{\partial}{\partial y_i}\left[D_0(\displaystyle\frac{\partial c^3}{\partial y_i}+\displaystyle\frac{\partial c^2}{\partial x_i}  )\right]=
   -D_0\gamma_{ijk}^1\displaystyle\frac{\partial^3 c^0}{\partial x_i\partial x_j\partial x_k} \vspace{0,2cm}\nonumber\\ 
   +(v_i^0\gamma_{ijk}^1-D_0\displaystyle\frac{\partial \pi_k}{\partial y_j})\displaystyle\frac{\partial^2 c^0}{\partial x_j\partial x_k}
   -D_0\gamma_{ij}^0\displaystyle\frac{\partial^2 \bar c^1}{\partial x_i\partial x_j} \label{CD_eq_defc2_3}\vspace{0,2cm}\\
    +(v_i^0\displaystyle\frac{\partial \pi_j}{\partial y_i}+v_i^1\gamma_{ij}^0)\displaystyle\frac{\partial c^0}{\partial x_j}
     +v_i^0\gamma_{ij}^0\displaystyle\frac{\partial \bar c^1}{\partial x_j} 
     +\displaystyle\frac{\partial c^1}{\partial t}.\nonumber
\end{numcases}
We may now determine an expression for ${\partial c^1}/{\partial t}$. From the definition of $c^1$ (Eq. (\ref{D_defc1})), we have
\begin{equation}
 \begin{array}{l}
  \displaystyle\frac{\partial c^1}{\partial t}=
  \chi_i\displaystyle\frac{\partial}{\partial t}(\displaystyle\frac{\partial c^0}{\partial x_i})+\displaystyle\frac{\partial \bar c^1}{\partial t}=
   \chi_i\displaystyle\frac{\partial}{\partial x_i}(\displaystyle\frac{\partial c^0}{\partial t})+\displaystyle\frac{\partial \bar c^1}{\partial t}.
 \end{array}
\end{equation}
Now, using the expression of ${\partial c^0}/{\partial t}$ (Eq. (\ref{CD_partialc0partialt}))  and deducing ${\partial \bar c^1}/{\partial t}$ from
Eq. (\ref{2nd_order__corr_macro}), the above equation finally becomes
\begin{numcases}{}
  \displaystyle\frac{\partial c^1}{\partial t}= 
 \displaystyle\frac{1}{\phi} \chi_iD_{jk}\displaystyle\frac{\partial^3c^ 0}{\partial x_i \partial x_j \partial x_k}\vspace{0,2cm}\nonumber\\
 + \displaystyle\frac{1}{\phi} (D'_{ij}-\chi_i <v_j^0>)\displaystyle\frac{\partial^2 c^0}{\partial x_i \partial x_j}
 +\displaystyle\frac{1}{\phi}D_{ij}\displaystyle\frac{\partial^2 \bar c^1}{\partial x_i \partial x_j}\vspace{0,2cm} \label{partialc1partialt}\\
 - \displaystyle\frac{1}{\phi} (\chi_i \displaystyle\frac{\partial <v_j^0>}{\partial x_i}+<v_j^1>)\displaystyle\frac{\partial c^0}{\partial x_j}
 -\displaystyle\frac{1}{\phi} <v_i^0> \displaystyle\frac{\partial \bar c^1}{\partial x_i}.\nonumber
\end{numcases}
Then, from the expression obtained for $c^2$, Eq. (\ref{CD_defc2}), we get
\begin{equation}
 \displaystyle\frac{\partial c^2}{\partial x_i}= \eta_{jk}\displaystyle\frac{\partial^3 c^0}{\partial x_i\partial x_j \partial x_k}
 + \pi_j \displaystyle\frac{\partial^2 c^0}{\partial x_i \partial x_j}+\chi_j \displaystyle\frac{\partial^2 \bar c^1}{\partial x_i \partial x_j}+
 \displaystyle\frac{\partial \bar c^2}{\partial x_i}.
 \label{CD_gradxc2}
\end{equation}
Finally, substituting Eqs. (\ref{partialc1partialt}) and (\ref{CD_gradxc2})
into Eq. (\ref{CD_eq_defc2_3}), we get Eq. (\ref{defc3_eq1}), and the boundary condition Eq. (\ref{CD_bc_defc3}) over $\Gamma$  becomes Eq. (\ref{defc3_eq2}).
%
\subsection{Variational formulation of the local boundary value problem}
%
\label{c3_var_form}
The variational formulation of the the local problem defined by Eqs. (\ref{defc3_eq1}) and (\ref{defc3_eq2})  is obtained by
multiplying  both equations by a test function $\alpha$ of zero average and by integrating  over $\Omega_p$:
\begin{numcases}{}
 \displaystyle \int_{\Omega_p}\ \displaystyle\frac{\partial \alpha}{\partial y_i}[D_0(\displaystyle\frac{\partial c^3}{\partial y_i}
 +\eta_{jk}\displaystyle\frac{\partial^3 c^0}{\partial x_i \partial x_j \partial x_k}
  +\pi_j \displaystyle\frac{\partial^2 c^0}{\partial x_i \partial x_j}+\chi_j \displaystyle\frac{\partial^2 \bar c^1}{\partial x_i \partial x_j}
  +\displaystyle\frac{\partial \bar c^2}{\partial x_i})]\ d\Omega=\vspace{0,2cm}\nonumber\\
 - \displaystyle \int_{\Omega_p}\ \alpha (\displaystyle\frac{1}{\phi}\chi_i D_{jk} -D_0 \gamma_{ijk}^1)\ d\Omega
  \displaystyle\frac{\partial^3c^ 0}{\partial x_i \partial x_j \partial x_k}\vspace{0,2cm}\nonumber\\
  -\displaystyle \int_{\Omega_p}\ \alpha (v_i^0 \gamma_{ijk}^1- D_0 \displaystyle\frac{\partial \pi_k}{\partial y_j}-\displaystyle\frac{1}{\phi}\chi_j <v_k^0>)
  \ d\Omega  \displaystyle\frac{\partial^2 c^0}{\partial x_j \partial x_k}\vspace{0,2cm}\nonumber\\
  +\displaystyle \int_{\Omega_p}\ \alpha D_0\gamma_{ij}^0\ d\Omega \displaystyle\frac{\partial^2 \bar c^1}{\partial x_i \partial x_j}
  \vspace{0,2cm} \label{c3_var_form_eq}\\
  - \displaystyle \int_{\Omega_p}\ \alpha(v_i^0\displaystyle\frac{\partial \pi_j}{\partial y_i}+v_i^1 \gamma_{ij}^0
  - \displaystyle\frac{1}{\phi}\chi_i \displaystyle\frac{\partial <v_j^0>}{\partial x_i})
 \ d\Omega\displaystyle\frac{\partial c^0}{\partial x_j}\vspace{0,2cm}\nonumber\\
 -\displaystyle \int_{\Omega_p}\ \alpha v_i^0\gamma_{ij}^0\ d\Omega\displaystyle\frac{\partial \bar c^1}{\partial x_j}.\nonumber
\end{numcases}
%
\subsection{Definition of the third-order tensor $\xi_{lmp}$}
%
\label{def_xi}
$\xi_{lmp}$ is the solution for $c^3$ when 
$$
\displaystyle\frac{\partial^3 c^0}{\partial x_i\partial x_j \partial x_k}=\delta_{il}\delta_{jm}\delta_{kp},
$$
while the other forcing
terms are set to zero. Thus, $\xi_{lmp}$ must satisfy
\begin{equation}
\left\lbrace
 \begin{array}{l}
  \displaystyle\frac{\partial}{\partial y_i}\left[D_0(\displaystyle\frac{\partial \xi_{lmp}}{\partial y_i}+\eta_{mp} \delta_{il}) \right]=
  \displaystyle\frac{1}{\phi}\chi_l D_{mp}
  - D_0\gamma_{lmp}^1\hspace{0,5cm}\hbox{within $\Omega_p$},\vspace{0,2cm}\\
  \left[D_0(\displaystyle\frac{\partial \xi_{lmp}}{\partial y_i}+\eta_{mp} \delta_{il}) \right]\ n_i = 0
  \hspace{0,5cm}\hbox{over $\Gamma$},\vspace{0,2cm}\\
  <\xi_{lmp}>=0, \vspace{0,2cm}\\
  \hbox{${\bar{\bar{\bar \xi}}}$: $\Omega$-periodic.}
 \end{array}
 \right.
\end{equation}
From Eq. (\ref{c3_var_form_eq}), we deduce the corresponding variational formulation:
\begin{equation}
\begin{array}{l}
  \displaystyle\int_{\Omega_p}\displaystyle\frac{\partial \alpha}{\partial y_i}
  \left[D_0(\displaystyle\frac{\partial \xi_{lmp}}{\partial y_i}+\eta_{mp} \delta_{il}) \right]\ d\Omega=
  -\displaystyle\int_{\Omega_p}\ \alpha(\displaystyle\frac{1}{\phi}\chi_l D_{mp}- D_0 \gamma^1_{lmp})\ d\Omega.
  \end{array}
 \label{form_var_xi}
\end{equation}
%
\subsection{Definition of the second-order tensor ${\tau_{lm}} $}
%
\label{def_tau}
$\tau_{lm}$ is the solution for $c^3$ when
$$
\displaystyle\frac{\partial^2 c^0}{\partial x_i\partial x_j} =\delta_{il} \delta_{jm}.
$$
Thus, it is the solution to
 \begin{equation}
\left\lbrace
 \begin{array}{l} 
  \displaystyle\frac{\partial}{\partial y_i}\left[D_0 (\displaystyle\frac{\partial \tau_{lm}}{\partial y_i}+\pi_m \delta_{il}) \right]=\vspace{0,2cm}\\
  v_i^0 \gamma_{ilm}^1- D_0 \displaystyle\frac{\partial \pi_m}{\partial y_l}+ \displaystyle\frac{1}{\phi} D'_{lm}-\displaystyle\frac{1}{\phi}\chi_l <v_m^0>
  \hspace{0,5cm}\hbox{within $\Omega_p$,}\vspace{0,2cm}\\
  \left[D_0 (\displaystyle\frac{\partial \tau_{lm}}{\partial y_i}+\pi_m \delta_{il}) \right]\ n_i=0
   \hspace{0,5cm}\hbox{over $\Gamma$},\vspace{0,2cm}\\
   <\tau_{lm}>=0, \vspace{0,2cm}\\
  \hbox{${\bar{\bar \tau}}$: $\Omega$-periodic,}
   \end{array}
 \right.
\end{equation}
and the associated variational formulation reads
\begin{equation}
 \begin{array}{l}
   \displaystyle\int_{\Omega_p}\displaystyle\frac{\partial \alpha}{\partial y_i}\left[D_0 (\displaystyle\frac{\partial \tau_{lm}}{\partial y_i}+\pi_m \delta_{il}) \right]\ d\Omega =
   - \displaystyle\int_{\Omega_p}\alpha (v_i^0 \gamma_{ilm}^1- D_0 \displaystyle\frac{\partial \pi_m}{\partial y_l}-\displaystyle\frac{1}{\phi}\chi_l <v_m^0>)
   \ d\Omega.
   \label{var_form_tau}
 \end{array}
\end{equation}
%
\subsection{Definition of vector $\theta_k  $}
%
\label{def_theta}
$\theta_k $ is the solution for $c^3$  when
$$
\displaystyle\frac{\partial c^0}{\partial x_j}=\delta_{jk}.
$$
Therefore, it must satisfy
\begin{equation}
 \left\lbrace
 \begin{array}{l}
  \displaystyle\frac{\partial}{\partial y_i}(D_0\displaystyle\frac{\partial\theta_k}{\partial y_i})=
  v_i^0\displaystyle\frac{\partial \pi_k}{\partial y_i}+v_i^1 \gamma_{ik}^0
  - \displaystyle\frac{1}{\phi}\chi_i \displaystyle\frac{\partial <v_k^0>}{\partial x_i}-\displaystyle\frac{1}{\phi} <v_k^1>
   \hspace{0,5cm}\hbox{within $\Omega_p$,}\vspace{0,2cm}\\
   (D_0\displaystyle\frac{\partial\theta_k}{\partial y_i})\ n_i = 0
    \hspace{0,5cm}\hbox{over $\Gamma$},\vspace{0,2cm}\\
    <\theta_{k}>=0, \vspace{0,2cm}\\
  \hbox{$\overrightarrow{\eta}$: $\Omega$-periodic.}
 \end{array}
 \right.
\end{equation}
The corresponding variational formulation is
\begin{equation}
 \displaystyle\int_{\Omega_p}\displaystyle\frac{\partial \alpha}{\partial y_i}D_0\displaystyle\frac{\partial \theta_k}{\partial y_i}\ d\Omega =
 -\displaystyle\int_{\Omega_p}\alpha (v_i^0\displaystyle\frac{\partial \pi_k}{\partial y_i}+v_i^1 \gamma_{ik}^0
  - \displaystyle\frac{1}{\phi}\chi_i \displaystyle\frac{\partial <v_k^0>}{\partial x_i})\ d\Omega.
  \label{var_form_theta}
\end{equation}
%
\subsection{Properties of the second-order tensor $D''_{jk}$}
%
\label{Prop_D''}
By taking $\alpha=\theta_k$ in the variational formulation associated with
$\overrightarrow{\chi}$ (Eq. (\ref{form_var_chi})), we get:
\begin{equation}
 \int_{\Omega_{\mathrm p}}  \displaystyle\frac{\partial \theta_{k}}{\partial y_i}D_0\displaystyle\frac{\partial\chi_j}{\partial y_i}\ d\Omega= 
 - \int_{\Omega_{\mathrm p}} D_0 \displaystyle\frac{\partial \theta_{k}}{\partial y_j}\ d\Omega.
  \label{Prop_D-dble_prime_eq1}
\end{equation}
Next, we consider $\alpha=\chi_j$ in the variational formulation associated with $\overrightarrow\theta $ (Eq. (\ref{var_form_theta})):
\begin{equation}
 \displaystyle\int_{\Omega_p}\displaystyle\frac{\partial \chi_j}{\partial y_i}D_0\displaystyle\frac{\partial \theta_k}{\partial y_i}\ d\Omega =
 -\displaystyle\int_{\Omega_p}\chi_j (v_i^0\displaystyle\frac{\partial \pi_k}{\partial y_i}+v_i^1 \gamma_{ik}^0
  - \displaystyle\frac{1}{\phi}\chi_i \displaystyle\frac{\partial <v_k^0>}{\partial x_i})\ d\Omega.
  \label{Prop_D-dble_prime_eq2}
\end{equation}
Then, from Eqs. (\ref{Prop_D-dble_prime_eq1})-(\ref{Prop_D-dble_prime_eq2}), we deduce that:
\begin{equation}
 \int_{\Omega_{\mathrm p}} D_0 \displaystyle\frac{\partial \theta_{k}}{\partial y_j}\ d\Omega=\displaystyle\int_{\Omega_p}\chi_j (v_i^0\displaystyle\frac{\partial \pi_k}{\partial y_i}+v_i^1 \gamma_{ik}^0
  - \displaystyle\frac{1}{\phi}\chi_i \displaystyle\frac{\partial <v_k^0>}{\partial x_i})\ d\Omega.
\end{equation}
From the above relationship and from the expression of $D''_{jk}$, Eq. (\ref{def_D_dbleprime}), we get:
\begin{equation}
 D''_{jk}=<\chi_j (v_i^0\displaystyle\frac{\partial \pi_k}{\partial y_i}+v_i^1 \gamma_{ik}^0
  - \displaystyle\frac{1}{\phi}\chi_i \displaystyle\frac{\partial <v_k^0>}{\partial x_i}) > - < v_j^1\chi_k + v_j^0 \pi_k>,
\end{equation}
from which we see that $D''_{jk}$ can be determined from $\overrightarrow{\chi}$, $\overrightarrow{\pi}$, $\overrightarrow{v}^0$ and $\overrightarrow{v}^1$.
\subsection{Properties of the third-order tensor $E'_{jlm}$}
\label{Prop_E'}
Let us consider $\alpha=\tau_{lm} $ in the variational formulation associated with
$\overrightarrow{\chi}$ (Eq. (\ref{form_var_chi})):
\begin{equation}
 \int_{\Omega_{\mathrm p}}  \displaystyle\frac{\partial \tau_{lm}}{\partial y_i}D_0\frac{\partial\chi_j}{\partial y_i}\ d\Omega= 
 - \int_{\Omega_{\mathrm p}} D_0 \displaystyle\frac{\partial \tau_{lm}}{\partial y_j}\ d\Omega.
\end{equation}
We may now take $\alpha=\chi_j$ in the variational formulation associated with ${{\bar{\bar \tau}}} $ (Eq. (\ref{var_form_tau})):
\begin{equation}
\begin{array}{l}
 \displaystyle\int_{\Omega_{\mathrm p}}  \displaystyle\frac{\partial \tau_{lm}}{\partial y_i}D_0\frac{\partial\chi_j}{\partial y_i}\ d\Omega=\vspace{0,2cm} \\
 -\displaystyle\int_{\Omega_{\mathrm p}} \frac{\partial \chi_j}{\partial y_i} D_0 \pi_m \delta_{il}\ d\Omega\vspace{0,2cm} \\
 - \displaystyle\int_{\Omega_p}\chi_j (v_i^0\displaystyle\frac{\partial \pi_k}{\partial y_i}+v_i^1 \gamma_{ik}^0
  - \displaystyle\frac{1}{\phi}\chi_i \displaystyle\frac{\partial <v_k^0>}{\partial x_i})\ d\Omega.
 \end{array}
\end{equation}
From the above two equations, we get:
\begin{equation}
 \displaystyle\int_{\Omega_{\mathrm p}} D_0 \displaystyle\frac{\partial \tau_{lm}}{\partial y_j}\ d\Omega =
 \displaystyle\int_{\Omega_{\mathrm p}} D_0\frac{\partial \chi_l}{\partial y_j}  \pi_m \ d\Omega
 + \displaystyle\int_{\Omega_p}\chi_j (v_i^0\displaystyle\frac{\partial \pi_k}{\partial y_i}+v_i^1 \gamma_{ik}^0
  - \displaystyle\frac{1}{\phi}\chi_i \displaystyle\frac{\partial <v_k^0>}{\partial x_i})\ d\Omega.
\end{equation}
From the definition of $E'_{jlm}$, Eq. (\ref{def_E_prime}), and the above relationship, we finally obtain:
\begin{equation}
E'_{jlm}=<D_0\frac{\partial \chi_l}{\partial y_j}  \pi_m > +
<\chi_j (v_i^0\displaystyle\frac{\partial \pi_k}{\partial y_i}+v_i^1 \gamma_{ik}^0
  - \displaystyle\frac{1}{\phi}\chi_i \displaystyle\frac{\partial <v_k^0>}{\partial x_i}) >- < v_i^0\eta_{jk}>.
\end{equation}
Therefore, tensor $E'_{jlm}$ can be determined from $\overrightarrow{\chi}$, $\overrightarrow{\pi}$, ${\bar{\bar \eta}}$, $\overrightarrow{v}^0$ and $\overrightarrow{v}^1$.
\subsection{Properties of the fourth-order tensor $F_{jlmp}$}
%
\label{Prop_F}
Let firstly take $\alpha=\xi_{lmp} $ in the variational formulation associated with
$\overrightarrow{\chi}$ (Eq. (\ref{form_var_chi})):
\begin{equation}
 \int_{\Omega_{\mathrm p}} D_0 \displaystyle\frac{\partial \xi_{lmp}}{\partial y_i}\displaystyle\frac{\partial\chi_j}{\partial y_i}\ d\Omega= 
 - \int_{\Omega_{\mathrm p}} D_0 \displaystyle\frac{\partial \xi_{lmp}}{\partial y_j}\ d\Omega.
\end{equation}
Next, by considering $\alpha=\chi_j$ in the variational formulation associated with ${\bar{\bar{\bar \xi}}} $ (Eq. (\ref{form_var_xi})), we get:
\begin{equation}
\begin{array}{l}
 \displaystyle\int_{\Omega_{\mathrm p}} D_0 \displaystyle\frac{\partial \xi_{lmp}}{y_i}\displaystyle\frac{\partial\chi_j}{\partial y_i}\ d\Omega= 
 - \displaystyle\int_{\Omega_{\mathrm p}} D_0 \displaystyle\frac{\partial \chi_j}{\partial y_i}\eta_{mp}\delta_{il}\ d\Omega
 -\displaystyle\frac{1}{\phi}\displaystyle\int_{\Omega_{\mathrm p}} D_{mp}\chi_j\chi_l\ d\Omega
 +\displaystyle\int_{\Omega_{\mathrm p}} \chi_j\gamma_{lmp}^1\ d\Omega.
 \end{array}
\end{equation}
From the above two relationships, and from the definition of $ F_{jlmp} $ (Eq. (\ref{def_F})), we deduce that:
\begin{equation}
 F_{jlmp}=<D_0 \eta_{mp}\displaystyle\frac{\partial \chi_l}{\partial y_j} >- <\chi_j \gamma^1_{lmp}>
 +\displaystyle\frac{1}{\phi} <\chi_j \chi_l D_{mp}> +< \eta_{jl}\delta_{mp}>,
\end{equation}
which shows that $F_{jlmp}$ is determined from $\overrightarrow{\chi}$ and ${\bar{\bar \eta}}$.
\section{Macroscopic fluxes}
\setcounter{equation}{0}
\subsection{Derivation of $<q_i^1 >_{\Sigma_{p_i}} $}
\label{Det_surf_av_q1}
To determine $<q_i^1 >_{\Sigma_{p_i}} $, we see from Eq. (\ref{def_surf_mean_qi1}), that the term $<y_i{\partial q_j^1}/{\partial y_j}> $ 
must be determined. This can be done by starting from Eq. (\ref{partialq1partialy_1}). 
By Eq. (\ref{Flux_eq_1_1}), we firstly deduce that
\begin{equation}
 \displaystyle\frac{\partial q_j^0}{\partial x_j}=-D_0 \gamma_{jk}^0\displaystyle\frac{\partial^2 c^0}{\partial x_j\partial x_k}+v_j^0\displaystyle\frac{\partial c^0}{\partial x_j}
 +c^0\displaystyle\frac{\partial v_j^0}{\partial x_j},
 \label{partialq0partialx}
\end{equation}
and then from Eq. (\ref{CD_1st_macro_1}), we get
\begin{equation}
 \displaystyle\frac{\partial c^0}{\partial t}=\displaystyle\frac{1}{\phi}D_{jk}\displaystyle\frac{\partial^2 c^0}{\partial x_j\partial x_k}-
 \displaystyle\frac{1}{\phi}< v_j^0>\displaystyle\frac{\partial c^0}{\partial x_j}.
 \label{partialc0partialt}
\end{equation}
Reporting expressions Eqs. (\ref{partialq0partialx}) and (\ref{partialc0partialt}) into Eq. (\ref{partialq1partialy_1}), we get:
\begin{equation}
 \displaystyle\frac{\partial q_j^1}{\partial y_j}=
 (D_0\gamma_{jk}^0-\displaystyle\frac{1}{\phi}D_{jk})\displaystyle\frac{\partial^2 c^0}{\partial x_j\partial x_k}- 
 (v_j^0 - \displaystyle\frac{1}{\phi}<v_j^0 >)\displaystyle\frac{\partial c^0}{\partial x_j}-c^0\displaystyle\frac{\partial v_j^0}{\partial x_j},
\end{equation}
from which we deduce
\begin{equation}
 <y_i \displaystyle\frac{\partial q_j^1}{\partial y_j}>=
 <D_0 y_i\gamma_{jk}^0-\displaystyle\frac{1}{\phi}y_iD_{jk}>\displaystyle\frac{\partial^2 c^0}{\partial x_j\partial x_k}
 -<y_i(v_j^0 -\displaystyle\frac{1}{\phi}<v_j^0>)>\displaystyle\frac{\partial c^ 0}{\partial x_j}
 -c^0<y_i \displaystyle\frac{\partial v_j^0}{\partial x_j}>.
 \label{av_yipartialq1ipartialyj}
\end{equation}
Now, reporting the above expression together with Eq. (\ref{flux_2nd_ord_macro_moy_q1}) into Eq. (\ref{def_surf_mean_qi1}),
we obtain the following expression for the surface average of $q_i^1$:
\begin{numcases}{}
 <q_i^1>_{\Sigma_{\mathrm p_i}}=
  - (E_{ijk}-<D_0 y_i\gamma_{jk}^0 - \displaystyle\frac{1}{\phi}y_iD_{jk} >)\displaystyle\frac{\partial^2 c^0}{\partial x_j\partial x_k}\vspace{0,2cm}\nonumber\\
  -(D'_{ij}-<y_i(\displaystyle\frac{1}{\phi}<v_j^0>-v_j^0  >)\displaystyle\frac{\partial c^0}{\partial x_j}- D_{ij}\displaystyle\frac{\partial \bar c^1}{\partial x_j}
  \vspace{0,2cm}\label{surf_mean_q1_0}\\
  -c^0<y_i\displaystyle\frac{\partial v_j^0}{\partial x_j}> 
  +c^0<v_i^1> + \bar c^1 <v_i^0>.\nonumber
\end{numcases}
To be physically meaningful, the macroscopic fluid velocity must also be defined by a surface average.
In order to determine $<v_i^1>_{\Sigma_{\mathrm p_i}}$, let consider the identity
\begin{equation}
 \displaystyle\frac{\partial}{\partial y_j}(y_i v_j^1)= y_i\displaystyle\frac{\partial v_j^1}{\partial y_j}+v_j^1.
\end{equation}
Integrating over $\Omega_\mathrm p$, we get
\begin{equation}
 <v_i^1>_{\Sigma_{p_i}}=<y_i\displaystyle\frac{\partial v_j^1}{\partial y_j} >+<v_i^1>.
\end{equation}
Now,  since by Eq. (\ref{F_eq2})  at ${\mathcal O}(\varepsilon^0)$
\begin{equation}
 \displaystyle\frac{\partial v_j^1}{\partial y_j}=-\displaystyle\frac{\partial v_j^0}{\partial x_j},
\end{equation}
we deduce that
\begin{equation}
 <v_i^1>_{\Sigma_{p_i}}=-<y_i\displaystyle\frac{\partial v_j^0}{\partial x_j} >+<v_i^1>.
 \label{av_v1Sigma_vs_av_v1Omega}
\end{equation}
Substituting Eq. (\ref{av_v1Sigma_vs_av_v1Omega}) into Eq. (\ref{surf_mean_q1_0}) and bearing in mind Eq. (\ref{v0_volmean_surf_mean}),
we finally get Eq. (\ref{q1surface_mean_final}) with Eqs. (\ref{def_ESigma})-(\ref{def_D'Sigma}).
\subsection{Derivation of $<q_i^2 >_{\Sigma_{p_i}} $}
\label{Det_surf_av_q2}
Let consider Eq. (\ref{identity}) with $q_i=q_i^2$
and integrate over $\Omega_\mathrm p$:
\begin{equation}
 <q_i^2>_{\Sigma_{\mathrm p_i}}=<y_i\displaystyle\frac{\partial q_j^2}{\partial y_j}>+<q_i^2>.
 \label{def_surf_mean_qi2}
\end{equation}
From Eq. (\ref{Flux_eq_2_3}), we get
\begin{equation}
 \displaystyle\frac{\partial q_j^2}{\partial y_j}=-\displaystyle\frac{\partial q_j^1}{\partial x_j}-\displaystyle\frac{\partial c^1}{\partial t}.
 \label{partialq2partialy_1}
\end{equation}
Using the definition of $ \overrightarrow{q}^1$ (Eq. (\ref{Flux_eq_1_2})), we deduce:
\begin{numcases}{}
  \displaystyle\frac{\partial q_j^1}{\partial x_j}= - D_0 \gamma_{jkl}^1\displaystyle\frac{\partial^3 c^0}{\partial x_j \partial x_k \partial x_l}
  - (D_0 \displaystyle\frac{\partial \pi_k}{\partial y_j}-v_j^0 \chi_k) \displaystyle\frac{\partial^2 c^0}{\partial x_j\partial x_k}\vspace{0,2cm}\nonumber\\
  - D_0 \gamma_{jk}^0 \displaystyle\frac{\partial^2 \bar c^1}{\partial x_j \partial x_k}
  + c^0 \displaystyle\frac{\partial v_j^1}{\partial x_j}+v_j^1 \displaystyle\frac{\partial c^0}{\partial x_j}
  +\bar c^1\displaystyle\frac{\partial v_j^0}{\partial x_j}+v_j^0 \displaystyle\frac{\partial \bar c^1}{\partial x_j},
\end{numcases}
and by Eq. (\ref{2nd_order__corr_macro}), we get
\begin{equation}
 \displaystyle\frac{\partial \bar c^1}{\partial t}= \displaystyle\frac{1}{\phi} D'_{jk}\displaystyle\frac{\partial^2 c^0}{\partial x_j\partial x_k}
 + \displaystyle\frac{1}{\phi} D_{jk}\displaystyle\frac{\partial^2 \bar c^1}{\partial x_j\partial x_k}
 - \displaystyle\frac{1}{\phi} <v_j^1>\displaystyle\frac{\partial c^0}{\partial x_j}-\displaystyle\frac{1}{\phi} <v_j^0>\displaystyle\frac{\partial \bar c^1}{\partial x_j}.
 \label{partial_bar_c1_patial_t}
\end{equation}
Using Eqs. (\ref{partialq2partialy_1})-(\ref{partial_bar_c1_patial_t}), we  deduce
\begin{numcases}{}
  <y_i \displaystyle\frac{\partial q_j^2}{\partial y_j}   > =  < D_0 y_i \gamma_{jkl}^1>\displaystyle\frac{\partial^3 c^0}{\partial x_j\partial x_k \partial x_l}\vspace{0,2cm}\nonumber\\
  + <D_0y_i \displaystyle\frac{\partial \pi_k}{\partial y_j}- y_iv_j^0 \chi_k - \displaystyle\frac{1}{\phi}y_i D'_{jk} >\displaystyle\frac{\partial^2 c^0}{\partial x_j \partial x_k}
  \vspace{0,2cm}\nonumber\\
  +< D_0 y_i\gamma_{jk}^0 - \displaystyle\frac{1}{\phi}y_i D_{jk}>\displaystyle\frac{\partial^2 \bar c^1}{\partial x_j\partial x_k}\vspace{0,2cm}\nonumber\\
  - <y_i (v_j^1 - \displaystyle\frac{1}{\phi} < v_j^1>) >\displaystyle\frac{\partial c^0}{\partial x_j}\vspace{0,2cm}\nonumber\\
  - <y_i (v_j^0 - \displaystyle\frac{1}{\phi} < v_j^0>) >\displaystyle\frac{\partial \bar c^1}{\partial x_j}\vspace{0,2cm} \label{av_ydivq2}\\
   - c^0 <y_i \displaystyle\frac{\partial v_j^1}{\partial x_j} >-{\bar c}^1 <y_i \displaystyle\frac{\partial v_j^0}{\partial x_j} >.\nonumber
\end{numcases}
Using Eq. (\ref{identity}) successively for $q_i= v_i^1 $ and $q_i=v_i^2$, and integrating both resulting equations over $\Omega_p$,
we can easily show that
\begin{eqnarray}
 - \bar c^1 <y_i\displaystyle\frac{\partial v_j^0}{\partial x_j} >= \bar c^1 (< v_i^1>_{\Sigma_{p_i}}- < v_i^1>),\\
 -  c^0 <y_i\displaystyle\frac{\partial v_j^1}{\partial x_j} >= c^0 (< v_i^2>_{\Sigma_{p_i}}- < v_i^2>).
\end{eqnarray}
By Eqs. (\ref{def_surf_mean_qi2}) and (\ref{av_ydivq2}), and using both above equations, together with the expression of $<q_i^2 > $, Eq. (\ref{flux_2nd_ord_macro_moy_q2}), yields

\begin{numcases}{}
  <q_i^2 >_{\Sigma_{p_i}}=- (F_{ijkl} -  F_{ijkl}^{\Sigma}) \displaystyle\frac{\partial^3 c^0}{\partial x_j\partial x_k \partial x_l}\vspace{0,2cm}\nonumber\\
  - (E'_{ijk}-E_{ijk}^{'\Sigma})\displaystyle\frac{\partial^2 c^0}{\partial x_j \partial x_k}\vspace{0,2cm}\nonumber\\
  -(E_{ijk} - E_{ijk}^{\Sigma})\displaystyle\frac{\partial^2 \bar c^1}{\partial x_j\partial x_k}\vspace{0,2cm}\nonumber\\
  - (D^{''}_{ij} - D^{''\Sigma}_{ij} )\displaystyle\frac{\partial c^0}{\partial x_j}\vspace{0,2cm}\label{3rdorder_corrector_macro_flux}\\
   - (D'_{ij}- D^{'\Sigma}_{ij})\displaystyle\frac{\partial \bar c^1}{\partial x_j}\vspace{0,2cm}\nonumber\\
   - D_{ij}\displaystyle\frac{\partial \bar c^2}{\partial x_j}\vspace{0,2cm}\nonumber\\
  + c^0 <v_i^2>_{\Sigma_{p_i}} + \bar c^1 < v_i^1>_{\Sigma_{p_i}}+\bar c^2 < v_i^0>_{\Sigma_{p_i}},\nonumber
\end{numcases}
in which
\begin{numcases}{}
 F_{ijkl}^{\Sigma}= < D_0 y_i \gamma_{jkl}^1>,
 \label{def_FSigma}\\
 E_{ijk}^{'\Sigma}=<D_0 y_i \displaystyle\frac{\partial \pi_k}{\partial y_j}-y_i v_j^0\chi_k - \displaystyle\frac{1}{\phi}y_iD'_{jk} >,
 \label{E'Sigma}\\
 D^{''\Sigma}_{ij}= <y_i (\displaystyle\frac{1}{\phi}< v_j^1>-v_j^1) >.
 \label{D''Sigma}
\end{numcases}
From Eqs. (\ref{CD_3rd_macro_1}) and (\ref{def_surf_mean_qi2}), we get the following writing for the second corrector
of the macroscopic mass-balance equation
\begin{equation}
 \phi \displaystyle\frac{\partial \bar c^2}{\partial t}+\displaystyle\frac{\partial}{\partial x_i} <q_i^2 >_{\Sigma_{p_i}}=
 \displaystyle\frac{\partial}{\partial x_i}(< y_i \displaystyle\frac{\partial q_j^2}{\partial y_j}>)
\end{equation}
Then, using the expression of $< y_i \displaystyle\frac{\partial q_j^2}{\partial y_j}> $, Eq. (\ref{av_ydivq2}), it becomes
\begin{numcases}{}
 \phi \displaystyle\frac{\partial \bar c^2}{\partial t}+\displaystyle\frac{\partial}{\partial x_i} <q_i^2 >_{\Sigma_{p_i}}=\vspace{0,2cm}\nonumber\\
 \displaystyle\frac{\partial}{\partial x_i}[F_{ijkl}^{\Sigma}\displaystyle\frac{\partial^3 c^0}{\partial x_j\partial x_k\partial x_l}
 + E^{'\Sigma}_{ijk}\displaystyle\frac{\partial^2 c^0}{\partial x_j\partial x_k} + E^{\Sigma}_{ijk}\displaystyle\frac{\partial^2 \bar c^1 }{\partial x_j\partial x_k}]
 \vspace{0,2cm}\nonumber\\
 + \displaystyle\frac{\partial}{\partial x_i}[D^{''\Sigma}_{ij}\displaystyle\frac{\partial c^0}{\partial x_j}+ D^{'\Sigma}_{ij}\displaystyle\frac{\partial \bar c^1}{\partial x_j} ]
 \vspace{0,2cm} \label{3rdorder_corrector}\\
  -\displaystyle\frac{\partial}{\partial x_i}[ c^0 (<v_i^2 >-<v_i^2 >_{\Sigma_{p_i}} + c^1 (<v_i^1 >-<v_i^1 >_{\Sigma_{p_i}}).\nonumber
\end{numcases}
To obtain the expression of the third-order macroscopic description with respect to the macroscopic flux,
let firstly add Eq. (\ref{flux_1st_ord_macro_moy_q1}) 
to Eq. (\ref{2ndord_correc_avSigmaq1}) multiplied by $\varepsilon$ and to  Eq. (\ref{3rdorder_corrector}) multiplied by $\varepsilon^2$:
\begin{numcases}{}
  \phi\displaystyle\frac{\partial <c >}{\partial t}+\displaystyle\frac{\partial}{\partial x_i} (<q_i >_{\Sigma_{p_i}})=\vspace{0,2cm}\nonumber\\
  \displaystyle\frac{\partial}{\partial x_i}\left[ \varepsilon^2 F_{ijkl}^{\Sigma}\displaystyle\frac{\partial^3 <c>}{\partial x_j\partial x_k\partial x_l} 
  + (\varepsilon E_{ijk}^{\Sigma}+\varepsilon^2 E^{'\Sigma}_{ijk})\displaystyle\frac{\partial^2 <c>}{\partial x_j\partial x_k}
  \right.\vspace{0,2cm}\nonumber\\
  + \left.(\varepsilon D^{'\Sigma}_{ij}+\varepsilon^2 D^{''\Sigma}_{ij})\displaystyle\frac{\partial <c>}{\partial x_j}
  - <c> (< v_i>-< v_i>_{\Sigma_{p_i}})\right]\vspace{0,2cm}\label{3rd_order_desc_adim}\\
  + {\mathcal O}(\varepsilon^3) (\phi\displaystyle\frac{\partial <c >}{\partial t}),\nonumber
\end{numcases}
and then let add Eq. (\ref{flux_1st_ord_macro_moy_q1}) to Eq. (\ref{q1surface_mean_final}) multiplied $\varepsilon$ 
and to Eq. (\ref{3rdorder_corrector_macro_flux}) multiplied by $\varepsilon^2$
\begin{numcases}{}
  < q_i>_{\Sigma_{p_i}}= - \varepsilon^2 (F_{ijkl}-F^{\Sigma}_{ijkl})\displaystyle\frac{\partial^3 <c>}{\partial x_j\partial x_k\partial x_l}\vspace{0,2cm}\nonumber\\
  - [\varepsilon(E_{ijk}-E^{\Sigma}_{ijk}) + \varepsilon^2 (E'_{ijk}-E^{'\Sigma}_{ijk})]\displaystyle\frac{\partial^2 <c>}{\partial x_j\partial x_k}\vspace{0,2cm}\nonumber\\
  - [D_{ij}+\varepsilon (D'_{ij}- D^{'\Sigma}_{ij}) +\varepsilon^2 (D^{''}_{ij}-D^{''\Sigma}_{ij})]\displaystyle\frac{\partial <c>}{\partial x_j}
  \vspace{0,2cm} \label{3rd_order_macro_flux_adim}\\
   + <c><v_i>_{\Sigma_{p_i}}+{\mathcal O}(\varepsilon^3 < q_i>_{\Sigma_{p_i}})\nonumber,
\end{numcases}
where
\begin{numcases}{}
 <q_i>_{\Sigma_{p_i}} = <q_i^0>_{\Sigma_{p_i}} + \varepsilon <q_i^1>_{\Sigma_{p_i}} + \varepsilon^2 <q_i^2>_{\Sigma_{p_i}}
 +{\mathcal O}(\varepsilon^3 <q_i>_{\Sigma_{p_i}}),
 \\
<v_i>_{\Sigma_{p_i}} = <v_i^0>_{\Sigma_{p_i}} + \varepsilon <v_i^1>_{\Sigma_{p_i}} +  \varepsilon^2 <v_i^2>_{\Sigma_{p_i}}
+{\mathcal O}(\varepsilon^3 <v_i>_{\Sigma_{p_i}}). 
\end{numcases}
%

\end{document}